\documentclass[a4paper,12pt]{article}
\pdfoutput=1 

\usepackage{jheppub} 

\usepackage[T1]{fontenc} 
\usepackage{amsfonts}
\usepackage{color,xcolor}

\title{Free field realization of the Ding-Iohara algebra at general levels}

\author[a]{Zitao Chen}
\author[b]{and Xiang-Mao Ding}

\affiliation[a]{School of Mathematical Sciences, University of Chinese Academy of Sciences,\\
	Academy of Mathematics and Systems Science, Beijing 100190, China}
\affiliation[b]{State Key Laboratory of Mathematical Sciences (SKLMS), \\
	Academy of Mathematics and Systems Science, \\
	Chinese Academy of Sciences, Beijing 100190, China}

\emailAdd{chenzitao@amss.ac.cn}
\emailAdd{xmding@amss.ac.cn}

\abstract{
We present a unified free field realization of the Ding-Iohara algebra at arbitrary levels, which satisfies a generalized form of the Serre relations. This realization, constructed using six free boson fields, arises from a specialized factorization of the structure function in the defining relations of the algebra. Based on this construction, we then develop intertwining operators for the Ding-Iohara algebra.}


\begin{document} 
\maketitle
\flushbottom

\section{Introduction} 
The Ding-Iohara algebra was first introduced by J.-T. Ding and K. Iohara as a generalization of the quantum affine algebras \cite{DI97}. The algebra supplemented with Serre relations was subsequently studied by K. Miki as a two-parameter deformation of $W_{1+\infty}$ \cite{Mi07}. The resulting structure is known as the Ding-Iohara-Miki (DIM) algebra, which is also referred to as 
the quantum toroidal algebra of $\mathfrak{gl}_1$. This algebra has two central elements $C$ and $K$, and possesses an $SL(2,\mathbb{Z})$ symmetry. This paper focuses on the representation theory of the Ding-Iohara algebra. By definition, the DIM algebra constitutes a subclass of the Ding-Iohara algebra; thus, any representation of the DIM algebra automatically yields a representation of the Ding-Iohara algebra. To date, the most extensively studied representations, namely the vertical and horizontal representations, are all representations of the DIM algebra. Vertical representations, where one of the central elements $C$ of the DIM algebra acts as the identity, were studied in a series of works \cite{FFJMM1,FFJMM2,FJMM3}. Three classes of vertical representations have been defined: vector, Fock, and MacMahon representations, respectively. If we regard vertical representations as analogues of level-zero representations of quantum affine algebras correspondingly, then there is another family of representations called horizontal representations, referring to level-one representations. The horizontal representations are realized as vertex operators in \cite{FHHSY10}. The DIM algebra exhibits broad relevance in mathematics and physics. It connects to diverse algebraic structures, including the affine Yangian of $\mathfrak{gl}_1$ (see \cite{Tsy}), the shuffle algebra (see\cite{FT}) as well as quantum $\mathcal{W}$-algebras  \cite{AKOS,FR,SKAO}. Its applications in physics rely primarily on intertwining operators that connect vertical and horizontal representations. Notably, the Serre relations play no direct role in the construction or computation of these operators. In physics, it gives algebraic frameworks for refined topological vertices in topological string theory \cite{IKV}, and qq-characters in supersymmetric gauge theories via the Alday-Gaiotto-Tachikawa (AGT) duality \cite{Ne}. The AGT duality claims an equivalence between the instanton partition function 4d $\mathcal{N}=2$ gauge theories and the conformal blocks of 2D conformal field theories (CFTs) \cite{AGT}. Then the AGT duality provides an insight into the relationships between gauge theories and infinite-dimensional symmetries. This duality has been extended to a 5d (K-theoretic) setting \cite{AY}, and generalized into the broader BPS/CFT correspondence \cite{Ne}, where the qq-characters serve as gauge-theoretic analogs of the energy-momentum tensor (or the so-called T-operators). For a comprehensive review of the DIM algebra, see \cite{MNNZ}.

In topological string theory, the (refined) topological vertex \cite{AKMV,IKV} is an effective tool for computing string amplitudes of the topological A-model for toric Calabi-Yau threefolds. Graphically, the topological vertex is a trivalent vertex attached to each vertex in the dual toric diagram. The refined topological vertices are realized as intertwining operators of the DIM algebra in \cite{AFS}. Each leg of the trivalent vertex is associated with a horizontal representation or a vertical representation of the DIM algebra. Horizontal representations link to NS5-branes, whereas vertical representations associate with D5-branes. More generally, the DIM algebra representations of level $(l_1,l_2)$ correspond to $(l_2,l_1)$-branes in network matrix models \cite{AKMMMMOZ}. A 5d $\mathcal{N}=1$ supersymmetric gauge theory can be obtained from type IIB string theory. Consequently, the Nekrasov partition function can be derived from the refined topological vertex, which links the representation theory of the DIM algebra with gauge theories.

In supersymmetric gauge theory, there is an important class of physical observables, called qq-characters \cite{Ne}. The most significant feature of qq-characters is the regularity of their vacuum expectation values, which is related to the integrability of supersymmetric gauge theories. As observed in \cite{AKMMMMOZ,MMZ}, the underlying symmetry is the DIM algebra, and the qq-characters can be derived from its intertwiners. This was further studied in \cite{BFHMZ,BFMZZ}, etc.

Free field methods, a crucial tool for studying infinite-dimensional symmetries, were first used to construct affine Lie algebras \cite{Wak}. A notable example includes the level-one representation of the quantum affine algebra $U_q(\hat{\mathfrak{sl}}_2)$ \cite{JM}. Analogously, horizontal representations of the DIM algebra are analogous to level-one representations of quantum affine algebras, while free field realizations of horizontal representations were achieved in \cite{FHHSY10}. This paper generalizes the construction of horizontal representations. We develop a unified free field realization of the Ding-Iohara algebra at arbitrary levels, in which the Serre relations are superseded by a generalized form. Furthermore, we demonstrate that these Serre relations can be interpreted as constraints that enforce the cancellation of potential formal delta-function singularities within certain cubic relations. 
Previous free field realizations restricted the first level of the DIM algebra due to pole constraints in the structure function $S_3(z)$ (see Section 2). To overcome this limitation, we decompose specific currents into two terms, inspired by the coproduct structure, and adopt an alternative factorization of the structure function. Following approaches for quantum affine and elliptic quantum algebras \cite{CD, Ma}, higher-level representations necessitate multiple free bosonic fields.  We also need multiple free bosonic fields to meet our requirements of factorizations in the construction for general levels. 

Intertwining operators, critical for computing CFT correlation functions  \cite{FR92}, hold significance in AGT duality. Vector intertwiners acting on tensor products of vertical vector and free field horizontal representations were introduced in \cite{FHHSY10}, with applications to holomorphic blocks of 3d quiver gauge theories \cite{Ze}. Fock intertwiners \cite{AFS},  derived from vertical Fock representations, align with refined topological vertices and reproduce K-theoretic Nekrasov instanton partition functions  \cite{Ne03}. An $N$-folded version of the Fock intertwiner was constructed in \cite{BFHMZ}. Recent extensions replace vertical Fock representations with MacMahon representations \cite{AKMMSZ}. Motivated by these results, we construct intertwining operators using vector representations and our free field realizations, though their gauge-theoretic implications within AGT duality require further exploration.

The main purpose of this paper is to derive a unified free field realization of the 
Ding-Iohara algebra with arbitrary levels. We hope that our construction will provide new insights into the structure of the Ding-Iohara algebra. It will be shown that the previously known free field realizations can be embedded in ours in the case of horizontal representations. Since our free field realization generalizes earlier constructions, it may correspond to branes with arbitrary charges in string theory, extending earlier ideas in network matrix models. With an interest in AGT duality, we also construct vector intertwiners 
that are analogous to the ordinary ones. There are other attractive objects in this context, like the R-matrix and screening charges. This new free field realization provides a tool for studying these objects. 

The main content of the article is as follows. Section 2 defines the Ding-Iohara algebra and its representations. Next, Section 3 details the free field realization for arbitrary levels. Then, Section 4 constructs vector intertwiners and their duals. Finally, section 5 concludes with future perspectives.

\section{Ding-Iohara algebra}
\subsection{Notation}
Let $\mathcal{U}$ denote the Ding-Iohara algebra. There are three deformation parameters $q_1,q_2,q_3$ in this algebra, under the constraint 
\begin{equation}
q_1q_2q_3=1.
\end{equation}
We introduce a structure function $g(z)$:
\begin{equation}
g(z)=\frac{(1-q_1z)(1-q_2z)(1-q_3z)}{(1-q^{-1}_1z)(1-q^{-1}_2z)(1-q^{-1}_3z)}
\end{equation}
with two independent parameters $\gamma, \sigma$ by setting 
\begin{equation}
q_1=\gamma^{-1}\sigma,\quad 
q_2=\gamma^{-1}\sigma^{-1},\quad
q_3=\gamma^2.
\end{equation}
satisfying 
\begin{equation}
g(z^{-1})=g(z)^{-1}.
\end{equation}
For convenience, we also define the following functions with different 
factorizations of the structure function $g(z)$:
\begin{equation}
g_{+}(z)=\frac{(1-q_1z)(1-q_2z)(1-q_3z)}{1-z},
\end{equation}
and
\begin{equation}
g_{-}(z)=\frac{(1-q^{-1}_1z)(1-q^{-1}_2z)(1-q^{-1}_3z)}{1-z},
\end{equation}
then, we have 
\begin{equation}
g(z)=\frac{g_{+}(z)}{g_{-}(z)}.
\end{equation}
Furthermore, if we denote
\begin{equation}
S_{1}(z)=\frac{(1-q_2z)(1-q_3z)}{(1-z)(1-q^{-1}_1 z)},~~~~
S_{2}(z)=\frac{(1-q_1z)(1-q_3z)}{(1-z)(1-q^{-1}_2 z)},
\end{equation}
and 
\begin{equation}
S_{3}(z)=\frac{(1-q_1z)(1-q_2z)}{(1-z)(1-q^{-1}_3 z)},
\end{equation}
respectively, then we have
\begin{equation}
g(z)=\frac{S_1(z)}{S_1(q_1z)}=\frac{S_2(z)}{S_2(q_2z)}=\frac{S_3(z)}{S_3(q_3z)}.
\end{equation}
These factorizations of $g(z)$ are crucial in this paper. 

\subsection{Definition}
The Ding-Iohara algebra $\mathcal{U}$ can be defined in terms of the Drinfeld currents as follows: 
\begin{equation}
E(z)=\sum_{n\in \mathbb{Z}} E_{n} z^{-n}, \quad 
F(z)=\sum_{n\in \mathbb{Z}} F_{n} z^{-n}, \quad
K^{\pm}(z)=\sum_{n \geq 0} K^{\pm}_{n} z^{\mp n}. 
\end{equation}
Using these currents, $\mathcal{U}$ is defined by the following commutation relations:
\begin{equation} \label{re:KK}
[K^{\pm}(z),K^{\pm}(w)]=0,
\end{equation}
\begin{equation} \label{re:KpKn}
K^{+}(z)K^{-}(w)=\frac{g(Cw/z)}{g(C^{-1}w/z)} K^{-}(z)K^{+}(w),
\end{equation}
\begin{equation}\label{re:KpE}
K^{+}(z)E(w)=g(C^{-1/2} w/z)^{-1}E(w)K^{+}(z),
\end{equation}
\begin{equation}\label{re:KpF}
K^{+}(z)F(w)=g(C^{1/2} w/z)F(w)K^{+}(z),
\end{equation}
\begin{equation} \label{re:KnE}
E(w)K^{-}(z)=g(C^{-1/2} z/w)^{-1}K^{-}(z)E(w),
\end{equation}
\begin{equation}\label{re:KnF}
F(w)K^{-}(z)=g(C^{1/2} z/w)K^{-}(z)F(w),
\end{equation}
\begin{equation}\label{re:EE}
\begin{aligned}
&(z-q_1w)(z-q_2w)(z-q_3w)E(z)E(w)\\
&=(z-q^{-1}_1w)(z-q^{-1}_2w)(z-q^{-1}_3w)E(w)E(z),
\end{aligned}
\end{equation}
\begin{equation}\label{re:FF}
\begin{aligned}
&(z-q^{-1}_1w)(z-q^{-1}_2w)(z-q^{-1}_3w)F(z)F(w)\\
&=(z-q_1w)(z-q_2w)(z-q_3w)F(w)F(z).
\end{aligned}
\end{equation}
\begin{equation}\label{re:EF}
[E(z),F(w)]=\tilde{g}(\delta(Cw/z)K^{+}(C^{1/2}w)-\delta(C^{-1}w/z)K^{-}(C^{-1/2}w))
\end{equation}
Here $\tilde{g}$ is a normalized factor and we choose $\tilde{g}=(1-q_1)(1-q_2)(1-q_3)$. Delta function $\delta(z)$ is defined as a formal series   
\begin{equation}
\delta(z)=\sum_{n\in \mathbb{Z}} z^n.
\end{equation}
The algebra $\mathcal{U}$ is an associative algebra generated by $E_n$, $F_n(n\in\mathbb{Z})$, $K^{+}_n(n\geq 0)$, $K^{-}_n (n\leq 0)$ and $C$. By imposing $K:=K^{-}_0=(K^{+}_0)^{-1}$, there are essentially two central elements $C$ and $K$ in this algebra. By definition, this algebra is invariant in permuting the parameters $q_i (i=1,2,3)$. 

The DIM algebra, which we also denote by $\mathcal{U'}$ for convenience, is defined by the same generators as the Ding-Iohara algebra and satisfies relations (\ref{re:KK})--(\ref{re:EF}) together with the Serre relations:

\begin{equation} \label{re:Serre_E}
Sym_{z_1,z_2,z_3} \frac{z_2}{z_3}[E(z_1),[E(z_2),E(z_3)]]=0,
\end{equation}
\begin{equation} \label{re:Serre_F}
Sym_{z_1,z_2,z_3} \frac{z_2}{z_3}[F(z_1),[F(z_2),F(z_3)]]=0.
\end{equation}

\subsection{Representations of the Ding-Iohara algebra}
For a $\mathcal{U}$-module, we say that it has level $(\zeta_1, \zeta_2)$ 
if $(C, K)=(\zeta_1, \zeta_2)$, where $\zeta_1,\zeta_2\in\mathbb{C}^{\times}$. For a $\mathcal{U}'$-module, it's said to be a vertical representation if $C=1$, as well as a 
horizontal representation if $C=\gamma$. 

\subsubsection{Vertical representations}
Vertical representations have been systematically studied in \cite{FFJMM1,FFJMM2,FJMM3}. Three important classes of vertical representations have been constructed, which are called vector representations, Fock representations and MacMahon representations. When $C=1$, $K^{\pm}_n\ (n\in \mathbb{Z})$ commute with each other due to relation (\ref{re:KpKn}). Anyone of these representations has a basis that simultaneously diagonalizes $K^{\pm}_n$. 
\par
We focus on a vector representation $V_1(u)$ here, and the representation has a basis $\{[u]_i  | i\in\mathbb{Z}\}$ labeled by integers $i\in\mathbb{Z}$, and $u$ is a spectral parameter. The action of $\mathcal{U}'$ is defined as:
\begin{equation}
K^{+}(z)[u]_i=S_1(q^{i+1}_1u/z) [u]_i,
\end{equation}
\begin{equation}
K^{-}(z)[u]_i=S_1(q^{-i}_1z/u) [u]_i,
\end{equation}
\begin{equation}
E(z) [u]_i= \mathcal{E}\delta(q_1^{i+1}u/z)[u]_{i+1},
\end{equation}
\begin{equation}
F(z) [u]_i= \mathcal{F}\delta(q_1^{i}u/z)[u]_{i-1},
\end{equation}
where 
\begin{equation} \label{re:factor_in_vector}
\mathcal{E} \mathcal{F}=\tilde{g} \frac{(1-q_2^{-1})(1-q_3^{-1})}{1-q_1}
=(1-q_2)(1-q_3)(1-q_2^{-1})(1-q_3^{-1}).
\end{equation}
Since the constant part of $S_1(z)$ is 1, the second level of vector representation is $K=1$. Since the parameters $q_1, q_2, q_3$ are symmetric, one can similarly define $V_i(u)(i=2, 3)$ by using $S_i(z)\ (i=2,3)$ instead of $S_1(z)$. The Fock representation can be derived from the vector representation by semi-infinite construction and has a basis labeled by Young diagrams (see \cite{FFJMM1} for details).  Similarly, the MacMahon representation can be derived from the Fock representation by semi-infinite construction, and has a basis labeled by plane partitions. The second level of Fock representation is $K=q_i\ (i=1,2,3)$ while the second level of MacMahon representation can be an arbitrary $\zeta_2\in\mathbb{C}^{\times}$. 
\subsubsection{Horizontal representations}
We now recall the free field realization of horizontal representations. Introduce (q-)bosonic oscillators $a_n\ (n\in\mathbb{Z} \backslash \{ 0 \})$ satisfying commutation relations:
\begin{equation}
[a_n,a_m]=\frac{1}{n} (1-q_1^n)(1-q_2^n) \delta_{n+m,0} , \quad n>0 .
\end{equation}
Define vertex operators:
\begin{equation}
V^{+}(z)=\exp(-\sum_{n \geq 0} a_n z^{-n}),
\end{equation}
and 
\begin{equation}
V^{-}(z)=\exp(\sum_{n\geq 0} a_{-n} z^n).
\end{equation}
The fundamental operator product expansion (OPE) is 
\begin{equation}
V^{+}(z)V^{-}(w)=S_3(w/z)^{-1}:V^{+}(z)V^{-}(w):= S_3(w/z)^{-1}V^{-}(w)V^{+}(z)
\end{equation} 
where the normal ordering, denoted by $:\dots:$, means that all the positive modes $a_n\ (n>0)$ are placed on the right and all the negative modes $a_n\ (n<0)$ are placed on the left. The above OPE relation can be computed by use of the following identity:
\begin{equation}
\ln (1-z) =-\sum_{n > 0} \frac{z^n}{n} .
\end{equation}
and the Baker-Campbell-Hausdorff formula:
\begin{equation}
e^Ae^B=e^{[A,B]} e^Be^A
\end{equation}
if $[A,B]$ commutes with both $A$ and $B$. 
\par
\par
Then we can define a free field realization of horizontal representation $\mathcal{F}^{(\gamma,1)}$ with level $(\gamma,1)$ by 
\begin{equation}
E(z)\mapsto (1-q_3)(1-q_3^{-1})V^{-}(z)V^{+}(z) ,
\end{equation} 
\begin{equation}
F(z)\mapsto V^{-}(\gamma z)^{-1}V^{+}(\gamma^{-1} z)^{-1} ,
\end{equation}
\begin{equation}
K^{+}(z)\mapsto V^{+}(\gamma^{1/2}z)V^{+}(\gamma^{-3/2}z)^{-1} ,
\end{equation} 
\begin{equation}
K^{-}(z)\mapsto V^{-}(\gamma^{-1/2}z)V^{-}(\gamma^{3/2}z)^{-1}  .
\end{equation} 
Note that the positions of $q_1$ and $q_2$ are symmetric, while $q_3=\gamma^2$ is special in this construction. The factor $(1-q_3)(1-q_3^{-1})$ corresponds to the choice of the normalized factor $\tilde{g}$. 
\par
The checks of defining relations (\ref{re:KK})-(\ref{re:KnF}) are straightforward. We just present details of checking defining relations (\ref{re:EE})-(\ref{re:EF}) here. 
The Serre relations (\ref{re:Serre_E}) and (\ref{re:Serre_F}) can be derived through a direct, albeit intricate, operator product expansion (OPE) calculation. The details are provided in the Appendix. 

By virtue of the fundamental OPE relation, we obtain the OPE relations between $V^{-}(z)V^{+}(z)$ and $V^{-}(w)V^{+}(w)$:
\begin{equation}
V^{-}(z)V^{+}(z)V^{-}(w)V^{+}(w)=S_3(w/z)^{-1} V^{-}(z) V^{-}(w) V^{+}(z)V^{+}(w),
\end{equation}
\begin{equation}
V^{-}(w)V^{+}(w)V^{-}(z)V^{+}(z)=S_3(z/w)^{-1} V^{-}(z) V^{-}(w) V^{+}(z)V^{+}(w).
\end{equation}
Thus, the defining relation (\ref{re:EE}) is satisfied. In the sense of analytic continuation, this can be rewritten as 
\begin{align}
&V^{-}(z)V^{+}(z)V^{-}(w)V^{+}(w)
\\ \notag
&=  S_3(w/z)^{-1}S_3(z/w) V^{-}(w)V^{+}(w)V^{-}(z)V^{+}(z)
\\ \notag
&=  g(w/z)^{-1} V^{-}(w)V^{+}(w)V^{-}(z)V^{+}(z),
\end{align}
where structure function $g(z)$ is factorized as 
\begin{equation}
g(z)=\frac{S_3(z)}{S_3(q_3z)}.
\end{equation}
Similarly, $V^{-}(\gamma z)^{-1}V^{+}(\gamma^{-1} z)^{-1}$ and  $V^{-}(\gamma w)^{-1}V^{+}(\gamma^{-1} w)^{-1}$ obey relations
\begin{align}
&V^{-}(\gamma z)^{-1}V^{+}(\gamma^{-1} z)^{-1}V^{-}(\gamma w)^{-1}V^{+}(\gamma^{-1} w)^{-1}
\\ \notag
&=S_3(q_3w/z)^{-1} V^{-}(\gamma z)^{-1}V^{-}(\gamma w)^{-1} V^{+}(\gamma^{-1} z)^{-1} V^{+}(\gamma^{-1}w)^{-1},
\end{align}
and 
\begin{align}
&V^{-}(\gamma w)^{-1}V^{+}(\gamma^{-1} w)^{-1}V^{-}(\gamma z)^{-1}V^{+}(\gamma^{-1} z)^{-1}
\\ \notag
&=S_3(q_3z/w)^{-1} V^{-}(\gamma z)^{-1}V^{-}(\gamma w)^{-1} V^{+}(\gamma^{-1} z)^{-1} V^{+}(\gamma^{-1}w)^{-1}.
\end{align}
In the sense of analytic continuation, we have
\begin{align}
&V^{-}(\gamma z)^{-1}V^{+}(\gamma^{-1} z)^{-1}V^{-}(\gamma w)^{-1}V^{+}(\gamma^{-1} w)^{-1}
\\ \notag
&= S_3(q_3w/z)^{-1}S_3(q_3z/w) V^{-}(\gamma w)^{-1}V^{+}(\gamma^{-1}w)^{-1}V^{-}(\gamma z)^{-1}V^{+}(\gamma^{-1} z)^{-1}
\\ \notag
&= g(w/z) V^{-}(\gamma w)^{-1}V^{+}(\gamma^{-1}w)^{-1}V^{-}(\gamma z)^{-1}V^{+}(\gamma^{-1} z)^{-1},
\end{align}
which implies (\ref{re:FF}).
At last, let us check (\ref{re:EF}). Since
\begin{align}
&V^{-}(z)V^{+}(z)V^{-}(\gamma w)^{-1}V^{+}(\gamma^{-1} w)^{-1}
\\ \notag
&= S_3(\gamma w/z) V^{-}(z)V^{-}(\gamma w)^{-1}V^{+}(z)V^{+}(\gamma^{-1} w)^{-1},
\end{align}
and 
\begin{equation}
\begin{aligned}
&V^{-}(\gamma w)^{-1}V^{+}(\gamma^{-1} w)^{-1}V^{-}(z)V^{+}(z)\\
&= S_3(\gamma z/w) V^{-}(z)V^{-}(\gamma w)^{-1}V^{+}(z)V^{+}(\gamma^{-1} w)^{-1},
\end{aligned}
\end{equation}
we obtain the defining relation (\ref{re:EF}):
\begin{align}
&[V^{-}(z)V^{+}(z),V^{-}(\gamma w)^{-1}V^{+}(\gamma^{-1} w)^{-1}] \\ \notag
&=\frac{(1-q_1)(1-q_2)}{1-q_3^{-1}} (\delta(\gamma w/z)V^{+}(\gamma w)V^{+}(\gamma^{-1} w)^{-1}\\ \notag
&~~~~~~~~~~~~~~~~~~~~~~~~~-\delta(\gamma^{-1}w/z)V^{-}(\gamma^{-1}w)V^{-}(\gamma w)^{-1}).
\end{align}
The above equality holds due to the identity:
\begin{equation}
S_3(\gamma w/z)-S_3(\gamma z/w)=\frac{(1-q_1)(1-q_2)}{1-q_3^{-1}}(\delta(\gamma w/z)-\delta(\gamma^{-1}w/z)),
\end{equation}
and the following property of delta function
\begin{equation}
f(z) \delta(w/z) =f(w) \delta(w/z).
\end{equation}
\par
For $\zeta_2\in\mathbb{C}^{\times}$, we can introduce certain modification factors to obtain a horizontal representation $\mathcal{F}_u^{(\gamma,\zeta_2)}$ of level $(\gamma,\zeta_2)$ (cf. \cite{AKMMSZ}): 
\begin{equation}
E(z)\mapsto (1-q_3)(1-q_3^{-1})e(z) V^{-}(z)V^{+}(z),
\end{equation} 
\begin{equation}
F(z)\mapsto f(z) V^{-}(\gamma z)^{-1}V^{+}(\gamma^{-1} z)^{-1},
\end{equation}
\begin{equation}
K^{+}(z)\mapsto k^{+}(z) V^{+}(\gamma^{1/2}z)V^{+}(\gamma^{-3/2}z)^{-1},
\end{equation} 
\begin{equation}
K^{-}(z)\mapsto k^{-}(z) V^{-}(\gamma^{-1/2}z)V^{-}(\gamma^{3/2}z)^{-1}.
\end{equation} 
where the modification factors $e(z)$, $f(z)$, $k^{\pm}(z)$ satisfy the following relations:
\begin{equation}
k^{\pm}(z)=e(\gamma^{\pm 1/2}z) f(\gamma^{\mp 1/2}z),
\end{equation}  
and the constant parts of $k^{\pm}(z)$ are fixed by $\zeta_2$. Such as, when $\zeta_2=\gamma^k$, $k^{\pm}(z)$, $e(z)$ and $f(z)$ can be chosen as 
\begin{equation}
k^{\pm}(z)=\gamma^{\mp k}, \quad e(z)=\gamma^k z^{-k}, \quad f(z)=\gamma^{-k} z^k.
\end{equation} 
\subsubsection{Coproduct}
To consider the tensor product of two representations of $\mathcal{U}$, we shall use the formal coproduct:
\begin{equation} \label{re:coE}
\Delta(E(z))=E(z)\otimes 1 +K^{-}(C_1^{1/2}z)\otimes E(C_1z),
\end{equation} 
\begin{equation} \label{re:coF}
\Delta(F(z))=F(C_2z)\otimes K^{+}(C_2^{1/2}z)+1\otimes F(z),
\end{equation}
\begin{equation} \label{re:coKp}
\Delta(K^{+}(z)) = K^{+}(C_2^{1/2}z)\otimes K^{+}(C_1^{-1/2}z),
\end{equation}
\begin{equation} \label{re:coKn}
\Delta(K^{-}(z)) = K^{-}(C_2^{-1/2}z)\otimes K^{-}(C_1^{1/2}z),
\end{equation}
\begin{equation}
	\Delta(C)=C\otimes C,
\end{equation}
where $C_1=C\otimes1$ and $C_2=1\otimes C$. Though the right-hand sides contain infinite sums, we can use this coproduct whenever it is well-defined. It is easy to check that this coproduct can be used to obtaining tensor product of our construction below. And we will find similarity between the coproduct and our construction.
\section{New free field realization}
In this section, we present our new free field realization of $\mathcal{U}$ with arbitrary level $(\zeta_1,\zeta_2)\in \mathbb{C}^{\times} \times \mathbb{C}^{\times}$. Let us explain our idea briefly. As shown in the construction of horizontal representations above, the delta functions in the defining relation (\ref{re:EF}) originate from the equality 
\begin{equation}
S_3(\gamma z)-S_3(\gamma z^{-1})=\frac{(1-q_1)(1-q_2)}{1-q_3^{-1}}(\delta(\gamma z)-\delta(\gamma^{-1}z)).
\end{equation}
There are two poles $z=1,q_3$ in the rational function $S_3(z)$. The relative shift of these two poles determines that the relative shift of the two delta functions obtained above can only be $q_3$. This is in fact restricts $C^2=q_3$. Thus, only representations with special values $C=\pm \gamma$ would be available for the free field realization constructed in Section 2. To lift this restriction, we expect that the two delta functions can be produced separately. That is why we separate the current $E(z)$ into two terms like
\begin{equation}
*:\exp(A(z)):+\ *:\exp(B(z)):
\end{equation} 
and separate the current $F(z)$ into two terms like
\begin{equation}
*:\exp(C(z)):+\ *:\exp(D(z)):.
\end{equation} 
where $*$ stands for some sort of factors.
At the same time, we need another factorization of the structure function $g(z)$. Instead of 
\begin{equation}
g(z)=\frac{S_3(z)}{S_3(q_3z)},	
\end{equation}
we adopt 
\begin{equation}
g(z)=\frac{g_{+}(z)}{g_{-}(z)}.
\end{equation} 
Note that 
\begin{equation}
g_{+}(z^{-1})=z^{-2}g_{-}(z),
\end{equation}
which implies that nontrivial zero modes and conjugate momenta are needed in the construction. Similar to the case of constructing higher level representations of quantum affine algebras, we find that more than one boson is needed to meet our requirements in the construction.

We construct representations of $\mathcal{U}$ with level $(\zeta_1,1)$ for general $\zeta_1\in \mathbb{C}^{\times}$ first. Then representations of $\mathcal{U}$ with level $(\zeta_1,\zeta_2)$ can be obtained by introducing modification factors.

\subsection{Representations with level $(\zeta_1,1)$} 
To construct free field representations with level $(\zeta_1,1)$, we introduce six free bosons $a^i,\ b^i,\ c^i$ for $i=1,2$. The bosons $a^i (i=1,2)$ can be regarded as the Cartan part while  $b^i,\ c^i (i=1,2)$ are introduced as analogues of ghost fields. The commuting relations of $a^i$ are
\begin{equation}
    [a_n^1,a_m^1]=-\frac{1}{n} (1-q_1^n)(1-q_2^n)(1-q_3^n) \delta_{n+m,0} , \quad n>0,
\end{equation}
\begin{equation}
    [a_n^2,a_m^2]=\frac{1}{n} (1-q_1^n)(1-q_2^n)(1-q_3^n) \delta_{n+m,0} , \quad n>0.
\end{equation}
Denote
\begin{equation}
    X_{\pm}(z)=\sum_{n > 0} X_{\pm n} z^{\mp n}
\end{equation}
for $X=a^i,b^i,c^i$. Then we have 
\begin{equation} \label{ope:a1}
\exp(a_{+}^1(z)) \exp(a_{-}^1(w))=g(w/z)^{-1} \exp(a_{-}^1(w)) \exp(a_{+}^1(z)),	
\end{equation}
and 
\begin{equation} \label{ope:a2}
\exp(a_{+}^2(z)) \exp(a_{-}^2(w))=g(w/z) \exp(a_{-}^2(w)) \exp(a_{+}^2(z)),	
\end{equation}
when reversing the ordering of the positive mode part and the negative mode part.
The modes of $b^i$ obey 
\begin{equation}
[b_n^1,b_m^1]=[b_n^2,b_m^2]=
\frac{1}{n} (-1+q_1^n+q_2^n+q_3^n) \delta_{n+m,0} , \quad n>0,
\end{equation}
while the modes of $c^i$ satisfy
\begin{equation}
[c_n^1,c_m^1]=[c_n^2,c_m^2]=
\frac{1}{n} (-1+q_1^{-n}+q_2^{-n}+q_3^{-n}) \delta_{n+m,0} , \quad n>0.
\end{equation}
Thus we obtain
\begin{equation} \label{ope:b1}
\exp(b_{+}^1(z)) \exp(b_{-}^1(w))=g_{+}(w/z)^{-1} \exp(b_{-}^1(w)) \exp(b_{+}^1(z)),
\end{equation}
\begin{equation} \label{ope:b2}
\exp(b_{+}^2(z)) \exp(b_{-}^2(w))=g_{+}(w/z)^{-1} \exp(b_{-}^2(w)) \exp(b_{+}^2(z)),
\end{equation}
and
\begin{equation} \label{ope:c1}
\exp(c_{+}^1(z)) \exp(c_{-}^1(w))=g_{-}(w/z)^{-1} \exp(c_{-}^1(w)) \exp(c_{+}^1(z)),	
\end{equation}
\begin{equation} \label{ope:c2}
\exp(c_{+}^2(z)) \exp(c_{-}^2(w))=g_{-}(w/z)^{-1} \exp(c_{-}^2(w)) \exp(c_{+}^2(z))	.
\end{equation}
The bosons $b^i$ play important roles in the construction of the current $E(z)$, while the bosons $c^i$ are significant for the current $F(z)$. Since we need nontrivial zero mode part in the construction of $E(z)$ and $F(z)$, we introduce zero modes and conjugate momenta for $b^i$ and $c^i$ symmetrically:
\begin{equation}
[p_b^1,q_b^1]=[p_b^2,q_b^2]=1,
\end{equation}
\begin{equation}
[p_c^1,q_c^1]=[p_c^2,q_c^2]=1.
\end{equation}
Define 
\begin{equation}
b^1(z)=\sum_{n \neq 0} b^1_n z^{-n} + p_b^1(-\ln z +\frac{1}{2} \ln \zeta_1)+q_b^1,
\end{equation}
\begin{equation}
b^2(z)=\sum_{n \neq 0} b^2_n z^{-n} + p_b^2(-\ln z -\frac{1}{2} \ln \zeta_1)+q_b^2,
\end{equation}
and
\begin{equation}
c^1(z)=\sum_{n \neq 0} c^1_n z^{-n} + p_c^1(-\ln z +\frac{1}{2} \ln \zeta_1)+q_c^1,
\end{equation}
\begin{equation}
c^2(z)=\sum_{n \neq 0} c^2_n z^{-n} + p_c^2(-\ln z -\frac{1}{2} \ln \zeta_1)+q_c^2.
\end{equation}
Also denote the zero mode parts of these fields by $X_0(z)$ for $X=b^i,c^i$, such as
\begin{equation}
    b_0^1(z)= p_b^1(-\ln z +\frac{1}{2} \ln \zeta_1)+q_b^1.
\end{equation}

Consider the Fock space $\mathcal{F}_{abc}$ generated by the vacuum state $|\boldsymbol{0} \rangle $:
\begin{equation}
\mathcal{F}_{abc}=\mathbb{C} \{ a_n^{i},\ b_n^{i}, c_n^{i} | n < 0,\ i=1,2 \} 
|\boldsymbol{0} \rangle
\end{equation}
with the annihilation condition:
\begin{equation}
p_a^{i} |\boldsymbol{0} \rangle =
p_b^{i} |\boldsymbol{0} \rangle =
p_c^{i} |\boldsymbol{0} \rangle =0,
\qquad \ i=1,\ 2
\end{equation}
and
\begin{equation}
a_n^{i} |\boldsymbol{0} \rangle =
b_n^{i} |\boldsymbol{0} \rangle =
c_n^{i} |\boldsymbol{0} \rangle =0,
\qquad n > 0,\ i=1,\ 2.
\end{equation}
Define 
\begin{equation}
V=\mathcal{F}_{abc} \otimes 
(\bigoplus_{m,n\in\mathbb{Z}} \mathbb{C}e^{m(q_b^1-q_c^2)+n(q_b^2-q_c^1)}) 
\end{equation}
and $\mathcal{U}$ will be realized as actions on $V$.
\par
With the help of these free fields, we can construct representations of $\mathcal{U}$ with level $(\zeta_1,1)$. Define 
\begin{equation}
\varphi^{+}(z)=\exp(
a_{+}^1(\zeta_1^{1/2}z)
+a_{+}^2(\zeta_1^{-1/2}z)
+b_{+}^1(\zeta_1^{1/2}z)
+c_{+}^2(\zeta_1^{-1/2}z)
),
\end{equation}
\begin{equation}
\varphi^{-}(z)=\exp(
a_{-}^1(\zeta_1^{-1/2}z)
+a_{-}^2(\zeta_1^{1/2}z)
+b_{-}^2(\zeta_1^{-1/2}z)
+c_{-}^1(\zeta_1^{1/2}z)
) ,
\end{equation}
and 
\begin{align} \label{def:eta}
\eta (z)=&z^{-1}[
:\exp(a_{+}^1(z)+b^1(z)-c_{-}^2(\zeta_1^{-1} z)-c_{0}^2(\zeta_1^{-1} z)):
\\ \notag
&+:\exp(a_{-}^1(z)+b^2(z)-c_{+}^1(\zeta_1 z)-c_{0}^1(\zeta_1 z)):
],
\end{align}
\begin{align} \label{def:xi}
\xi (z)=&z^{-1}[
:\exp(a_{-}^2(z)+c^1(z)-b_{+}^2(\zeta_1^{-1} z)-b_{0}^2(\zeta_1^{-1} z)):
\\ \notag
&+:\exp(a_{+}^2(z)+c^2(z)-b_{-}^1(\zeta_1 z)-b_{0}^1(\zeta_1 z)):
].
\end{align}
As before the normal ordering, denoted by $:\dots:$, means that we shall put the positive modes while putting the negative modes on the left. Since there are nontrivial zero mode parts in the expressions, we shall put $p$'s on the right and put $q$'s on the right at the same time, such as 
\begin{equation}
	:\exp(a^1(z)):=\exp(a_{-}^1(z))e^{q_a^1}z^{p_a^1}\exp(a_{+}^1(z)).
\end{equation}
Then the following homomorphism gives a representation of $\mathcal{U}$ with level $(\zeta_1,1)$:
\begin{equation}
	E(z) \mapsto \eta(z),\quad 
	F(z) \mapsto \xi(z) ,\quad
	K^{+}(z) \mapsto \varphi^{+}(z),\quad
	K^{-}(z) \mapsto \varphi^{-}(z) .
\end{equation}
\par
Although the proof is just a straightforward calculation of operator product expansion (OPE), our construction seems quite technical. We present the details of checking relations (\ref{re:KpKn}), (\ref{re:KpE}), (\ref{re:KnE}), (\ref{re:EE}) and (\ref{re:EF}), elucidating why this construction is effective. For the rest, relation (\ref{re:KK}) is obvious. The relations between $K^{\pm}(z)$ and $F(w)$ can be checked in the same way as those between $K^{\pm}(z)$ and $E(w)$, while (\ref{re:FF}) can be checked in the same way as (\ref{re:EE}). The Serre relations (\ref{re:Serre_E}) and (\ref{re:Serre_F}) will be discussed at a later stage. For simplicity, denote 
\begin{equation}
A(z)=a_{+}^1(z)+b^1(z)-c_{-}^2(\zeta_1^{-1} z)-c_{0}^2(\zeta_1^{-1}z),
\end{equation}
\begin{equation}
B(z)=a_{-}^1(z)+b^2(z)-c_{+}^1(\zeta_1 z)-c_{0}^1(\zeta_1z),
\end{equation}
\begin{equation}
C(z)=a_{-}^2(z)+c^1(z)-b_{+}^2(\zeta_1^{-1} z)-b_{0}^2(\zeta_1^{-1} z),
\end{equation}
\begin{equation}
D(z)=a_{+}^2(z)+c^2(z)-b_{-}^1(\zeta_1 z)-b_{0}^1(\zeta_1 z)
\end{equation}
so that 
\begin{equation}
\eta(z)=z^{-1} (:\exp(A(z)):+:\exp(B(z)):),
\end{equation}
\begin{equation}
\xi(z)=z^{-1} (:\exp(C(z)):+:\exp(D(z)):).
\end{equation}
\par
First, we prove the relation between $K^{\pm}(z)$ (\ref{re:KpKn}). Note that the only nontrivial commuting relation are those of $a^i\ (i=1,2)$ when reversing $\varphi^{+}(z)$ and $\varphi^{-}(w)$, which explains why we consider $a^i\ (i=1,2)$ as the Cartan part. Using (\ref{ope:a1}) and (\ref{ope:a2}), we get
\begin{equation}
\varphi^{+}(z) \varphi^{-}(w) =\frac{g(\zeta_1w/z)}{g(\zeta_1^{-1}w/z)} \varphi^{-}(w) \varphi^{+}(z).
\end{equation}
Oscillators $b^1_{n},c^2_{n},b^2_{-n},c^1_{-n}$ have no contribution in calculating OPE of $\varphi^{+}(z) \varphi^{-}(w)$, justifying their roles as analogues of ghost fields.
\par
Second, we consider relations between $\varphi^{\pm}(z)$ and $\eta(w)$, i.e. (\ref{re:KpE}) and (\ref{re:KnE}). Recall the defining relation (\ref{re:coE}) where $\Delta(E(z))$ is separated into two terms satisfying the same relations with $\Delta(K^{\pm}(z))$. The situation for $\eta(w)$ is similar. When computing the OPE of $\varphi^{+}(z):\exp(A(w)):$, nontrivial contributions come from $b^1$ and $c^2$. Using (\ref{ope:b1}) and (\ref{ope:c2}), we obtain
\begin{align}
\varphi^{+}(z) :\exp(A(w)):
&=\frac{g_{-}(\zeta_1^{-1/2}w/z)}{g_{+}(\zeta_1^{-1/2}w/z)} :\exp(A(w)):\varphi^{+}(z)
\\ \notag
&=g(\zeta_1^{-1/2}w/z)^{-1} :\exp(A(w)):\varphi^{+}(z).
\end{align}
In the computation of $\varphi^{+}(z) :\exp(B(w)):$, we only need to consider the effect of reversing the positive mode part and the negative mode part of $a^1$. Using (\ref{ope:a1}), we get
\begin{equation}
\varphi^{+}(z) :\exp(B(w)):= g(\zeta_1^{-1/2}w/z)^{-1}:\varphi^{+}(z) \exp(B(w)):.
\end{equation}
Thus 
\begin{equation}
\varphi^{+}(z) \eta(w)= g(\zeta_1^{-1/2}w/z)^{-1} \eta(w) \varphi^{+}(z) .
\end{equation}
We need to consider the effect of $a^1$ when calculating $:\exp(A(w)):\varphi^{-}(z)$. So we obtain 
\begin{equation}
    :\exp(A(w)):\varphi^{-}(z)=g(\zeta_1^{-1/2}z/w)^{-1}\varphi^{-}(z):\exp(A(w)):.
\end{equation}
using (\ref{ope:a1}). On the other hand, we shall take $b^2$ and $c^1$ into account when computing the OPE of $:\exp(B(w)):\varphi^{-}(z)$. Using (\ref{ope:b2}) and (\ref{ope:c1}), we have
\begin{align}
    :\exp(B(w)):\varphi^{-}(z)
    &=\frac{g_{-}(\zeta_1^{-1/2}z/w)}{g_{+}(\zeta_1^{-1/2}z/w)} \varphi^{-}(z):\exp(B(w)):
    \\ \notag
    &=g(\zeta_1^{-1/2}z/w)^{-1}\varphi^{-}(z):\exp(B(w)):.
\end{align}
Thus we obtain
\begin{equation}
    \eta(w)\varphi^{-}(z)=g(\zeta_1^{-1/2}z/w)^{-1}\varphi^{-}(z)\eta(w).
\end{equation}
In the free field realization of horizontal representations where only one boson is used, vertex operators are determined by their OPE relations between $K^{\pm}(z)$ uniquely up to a factor. However, we see that $:\exp(A(w)):$ and $:\exp(B(w)):$ are indeed two different vertex operators satisfying the same OPE relations with $\varphi^{\pm}(z)$, which is a distinguishing feature in the construction with multiple free bosons.
\par
Next, we consider the relation of $\eta(z)$ with itself (\ref{re:EE}). The structure of the coproduct inspires our construction, therefore, we also perform formal calculation for the coproduct here to illustrate our idea. To check (\ref{re:EE}) for $\Delta(E(z))$, we need to consider four terms 
\begin{align}
&\Delta(E(z))\Delta(E(w))
\\ \notag
&=E(z)E(w)\otimes 1 +K^{-}(C_1^{1/2}z)E(w)\otimes E(C_1z)
\\ \notag
&~~~+E(z)K^{-}(C_1^{1/2}w)\otimes E(C_1w) +K^{-}(C_1^{1/2}z)K^{-}(C_1^{1/2}w)\otimes E(C_1z)E(C_1w).
\end{align}
For the first term and the last term, the relation (\ref{re:EE}) is ensured by the defining relation of $E(z)$ in different representation spaces. However, the rest terms satisfy (\ref{re:EE}) due to the defining relation between $K^{-}(z)$ and $E(w)$. For $\eta(z)$, the boson $b^1$ plays the leading role in the relation of $:\exp(A(z)):$ with itself. Besides, it is necessary to take the zero-mode part into account. Combining (\ref{ope:b1}) with the following relations,  
\begin{equation}
[p_b^1(-\ln z+\frac{1}{2}\ln\zeta_1),q_b^1]=-\ln z+\frac{1}{2}\ln\zeta_1 ,
\end{equation}
\begin{equation}
[p_c^2(\ln z-\frac{1}{2}\ln\zeta_1),-q_c^2]=-\ln z+\frac{1}{2}\ln\zeta_1,
\end{equation}
we obtain
\begin{align}
&:\exp(A(z))::\exp(A(w)):\\
&= \frac{\zeta_1(z-w)}{(z-q_1w)(z-q_2w)(z-q_3w)}:\exp(A(z)+A(w)):
\end{align}
where the extra factor $\zeta_1$, which is not important here, is corresponded to the supports of delta functions in (\ref{re:EF}). Thus $:\exp(A(z)):$ and $:\exp(A(w)):$ satisfy
\begin{align}
&(z-q_1w)(z-q_2w)(z-q_3w):\exp(A(z))::\exp(A(w)):
\\ \notag
&=(z-q^{-1}_1w)(z-q^{-1}_2w)(z-q^{-1}_3w):\exp(A(w))::\exp(A(z)):.
\end{align}  
Similarly, the OPE of $:\exp(B(z))::\exp(B(w)):$ is given by 
\begin{align}
&:\exp(B(z))::\exp(B(w)):\\
&=\frac{\zeta_1^{-1} (z-w)}{(z-q_1w)(z-q_2w)(z-q_3w)}:\exp(B(z)+B(w)):
\end{align}
which is mainly obtained by reversing the positive mode part and the negative mode part of $b^2$. Thus 
\begin{align}
&(z-q_1w)(z-q_2w)(z-q_3w):\exp(B(z))::\exp(B(w)):
\\ \notag
&=(z-q^{-1}_1w)(z-q^{-1}_2w)(z-q^{-1}_3w):\exp(B(w))::\exp(B(z)):.
\end{align}  
Comparing with the coproduct, the bosons $b^i$ help us to realize the original defining relation of $E(z)$ with itself in different representation spaces. In the sense of analytic continuation, (\ref{re:EE}) can be written as
\begin{equation}
	E(z) E(w) = g(w/z)^{-1} E(w)E(z),
\end{equation}
and we factorize $g(w/z)^{-1}$ as 
\begin{equation}
	g(w/z)^{-1}=\frac{w^2 g_{+}(z/w)}{z^2 g_{+}(w/z)} =\frac{g_{-}(w/z)}{g_{+}(w/z)}.
\end{equation}
To compute the OPE of $:\exp(A(z))::\exp(B(w)):$, we only need to consider the effect of $a^1$, which is similar to that of $:\exp(A(w)):\varphi^{-}(z)$. Using (\ref{ope:a1}), we obtain
\begin{equation}
:\exp(A(z))::\exp(B(w)): = g(w/z)^{-1}:\exp(A(z)+B(w)):.
\end{equation}
And the OPE of $:\exp(B(w))::\exp(A(z)):$ is trivial:
\begin{equation}
	:\exp(B(w))::\exp(A(z)): = :\exp(A(z)+B(w)):.
\end{equation}
The OPE relations between $:\exp(A(z)):$ and $:\exp(B(w)):$ are the same as those between $:\exp(A(z)):$  and $\varphi^{-}(w)$,  except for a shift in mode. All these relations are essentially arising from the effect of the boson $a^{1}$.

Thus 
\begin{align}
&(z-q_1w)(z-q_2w)(z-q_3w):\exp(X(z))::\exp(Y(w)):
\\ \notag
& =(z-q^{-1}_1w)(z-q^{-1}_2w)(z-q^{-1}_3w):\exp(X(w))::\exp(Y(z)):,
\end{align}  
for $X=A,Y=B$ or $X=B,Y=A$. At this time, it can be regarded that we perform a trivial factorization to $g(w/z)^{-1}$. Checking the above four terms individually, we prove (\ref{re:EE}) for $\eta(z)$. Although the OPE factors of these four terms are different, they have the same total effects in the commuting relations, which we consider as different factorizations of the structure function $g(z)$.

\par
Last let us check the relation (\ref{re:EF}). From left hand side of (\ref{re:EF}), 
\begin{align}
[\eta(z),\xi(w)]=& \frac{1}{wz} ([:\exp(A(z)):,:\exp(C(w)):] 
\\ \notag
&~~~~+[:\exp(A(z)):,:\exp(D(w)):]
\\ \notag
&~~~~+[:\exp(B(z)):,:\exp(C(w)):]
\\ \notag
&~~~~ +[:\exp(B(z)):,:\exp(D(w)):]).
\end{align}
It is obvious that
\begin{equation} \label{eq:ACBD}
[:\exp(A(z)):,:\exp(C(w)):]=[:\exp(B(z)):,:\exp(D(w)):]=0.
\end{equation}
Commutators $[:\exp(A(z)):,:\exp(D(w)):]$ and $[:\exp(B(z)):,:\exp(C(w)):]$ correspond to $\varphi^{+}(z)$ and $\varphi^{-}(z)$ respectively. The OPE of $:\exp(A(z))::\exp(D(w)):$ is obtained by reversing the order of $b_{+}^1(z)$ and $b_{-}^1(\zeta_1w)$, $p_b^1$ and $q_b^1$, $p_c^2$ and $q_c^2$. Combining (\ref{ope:b1}) with 
\begin{equation}
[p_b^1(-\ln z+\frac{1}{2}\ln\zeta_1),-q_b^1]=\ln z-\frac{1}{2}\ln\zeta_1 ,
\end{equation}
\begin{equation}
[p_c^2(\ln z-\frac{1}{2}\ln\zeta_1),q_c^2]=\ln z-\frac{1}{2}\ln\zeta_1 ,
\end{equation}
we get
\begin{align} 
&:\exp(A(z))::\exp(D(w)):
\\ \notag 
&= \zeta_1^{-1} \frac{(z-q_1\zeta_1w)(z-q_2\zeta_1w)(z-q_3\zeta_1w)}{z-\zeta_1w} 
:\exp(A(z)+D(w)):
\end{align}
On the other hand, we need to reverse the order of $c_{+}^1(w)$ and $c_{-}^1(\zeta_1^{-1}z)$ when calculating the OPE of $:\exp(D(w))::\exp(A(z)):$. For the zero mode part, we also need to take $p_b^1$, $q_b^1$, $p_c^2$ and $q_c^2$ into account. Combining (\ref{ope:c2}) and 
\begin{equation}
[p_c^2(-\ln w-\frac{1}{2}\ln\zeta_1),-q_c^2]=\ln w+\frac{1}{2}\ln\zeta_1 ,
\end{equation}
\begin{equation}
[p_b^1(\ln w+\frac{1}{2}\ln\zeta_1),q_b^1]=\ln w+\frac{1}{2}\ln\zeta_1 ,
\end{equation}
we get
\begin{align} 
&:\exp(D(w))::\exp(A(z)):
\\ \notag
& = \zeta_1 \frac
{(w-q_1^{-1}\zeta_1^{-1}z)(w-q_2^{-1}\zeta_1^{-1}z)(w-q_3^{-1}\zeta_1^{-1}z)}
{w-\zeta_1^{-1}z} 
:\exp(A(z)+D(w)):.
\end{align}
Observing that 
\begin{equation}
\varphi^{+}(z)=: \exp(A(\zeta_1^{1/2} z)+D(\zeta_1^{-1/2}z)) :,
\end{equation}
we obtain 
\begin{align} \label{eq:AD}
&[:\exp(A(z)):,:\exp(D(w)):]
\\ \notag
&=(1-q_1)(1-q_2)(1-q_3) \zeta_1 w^2 \delta(\zeta_1w/z) \varphi^{+}(\zeta_1^{1/2} w) .
\end{align}
by using the following identity about delta function: 
\begin{equation} \label{eq:delta}
\frac{\prod_i 1-a_iz}{\prod_j 1-b_jz}-z^{|i|-|j|}\frac{\prod_i z^{-1}-a_i}{\prod_j z^{-1}-b_j}=\sum_k \frac{\prod_i 1-a_i/b_k}{\prod_{j\neq k} 1-b_j/b_k} \delta(b_k z).
\end{equation}
When computing the commutator $[:\exp(B(z)):,:\exp(C(w)):]$, we need to take $p_b^2$, $q_b^2$, $p_c^1$ and $q_c^1$ into account. $c^1$ plays the leading role in the computation of $:\exp(B(z))::\exp(C(w)):$ while it is $b^2$ that has significant effect in the computation of $:\exp(C(w))::\exp(B(z)):$. They obey
\begin{align} 
&:\exp(B(z))::\exp(C(w)): 
\\ \notag 
& = \zeta_1 \frac{(z-q_1^{-1}\zeta_1^{-1}w)(z-q_2^{-1}\zeta_1^{-1}w)(z-q_3^{-1}\zeta_1^{-1}w)}
{z-\zeta_1^{-1}w} :\exp(B(z)+C(w)):
\end{align}
and
\begin{align} 
& :\exp(C(w))::\exp(B(z)):
\\ \notag 
& = \zeta_1^{-1} \frac
{(w-q_1\zeta_1z)(w-q_2\zeta_1z)(w-q_3\zeta_1z)}
{w-\zeta_1z} :\exp(B(z)+C(w)):.
\end{align}
Since
\begin{equation}
\varphi^{-}(z)=:\exp(B(\zeta_1^{-1/2}z)+C(\zeta_1^{1/2}z)):,
\end{equation}
we get 
\begin{align} \label{eq:BC}
&[:\exp(B(z)):,:\exp(C(w)):]
\\ \notag
&=(1-q_1^{-1})(1-q_2^{-1})(1-q_3^{-1}) \zeta_1^{-1} w^2 \delta(\zeta_1^{-1}w/z) \varphi^{-}(\zeta_1^{-1/2} w).
\end{align}
 Combining (\ref{eq:ACBD}), (\ref{eq:AD}) and (\ref{eq:BC}), we find
\begin{equation}
[\eta(z),\xi(w)]=\tilde{g}(\delta(\zeta_1w/z)\varphi^{+}(\zeta_1^{\frac{1}{2}}w)
-\delta(\zeta_1^{-1}w/z)\varphi^{-}(\zeta_1^{-\frac{1}{2}}w)),
\end{equation}
concluding the relation (\ref{re:EF}). 

Instructively, we employ a special factorization of the structure function $g(z)$: 
\begin{equation}
g(z)=\frac{g_{+}(z)}{g_{-}(z)},
\end{equation}
to derive the representation of $\mathcal{U}$. In fact, if we assume that $\eta(z)$, $\xi(z)$ are written as
\begin{equation}
\eta(z)=\frac{1}{z}(:\exp(A(z)):+:\exp(B(z)):),
\end{equation}
\begin{equation}
\xi(z)=\frac{1}{z}(:\exp(C(z)):+:\exp(D(z)):),
\end{equation}
and $\varphi^{\pm}(z)$ are expressed as 
\begin{equation}\label{re:phi^p}
\varphi^{+}(z)=\exp(A_{+}(\zeta_1^{1/2}z)+D_{+}(\zeta_1^{-1/2}z)) ,
\end{equation}
\begin{equation}\label{re:phi^n}
\varphi^{-}(z)=\exp(B_{-}(\zeta_1^{-1/2}z)+C_{-}(\zeta_1^{1/2}z)),
\end{equation}
any realization of $A(z)$, $B(z)$, $C(z)$ and $D(z)$ with the following relations will lead to a representation of $\mathcal{U}$. First, the nonzero parts should obey 
\begin{equation}\label{re:AnDn}
D_{-}(z)=-A_{-}(\zeta_1 z),
\end{equation}
\begin{equation}\label{re:BpCp}
C_{+}(z)=-B_{+}(\zeta_1^{-1} z),
\end{equation}
\begin{equation}
[A_{+}(z),A_{-}(w)]=[B_{+}(z),B_{-}(w)]=-\ln g_{+}(w/z),
\end{equation}
\begin{equation}
[C_{+}(z),C_{-}(w)]=[D_{+}(z),D_{-}(w)]=-\ln g_{-}(w/z),
\end{equation}
\begin{equation}
[A_{+}(z),B_{-}(w)]=-\ln g(w/z),
\end{equation}
\begin{equation}
[D_{+}(z),C_{-}(w)]=\ln g(w/z),
\end{equation}
\begin{equation}\label{re:BpAn}
[B_{+}(z),A_{-}(w)]=[A_{+}(z),C_{-}(w)]=[D_{+}(z),B_{-}(w)]=0.
\end{equation}
Second, the zero mode part should satisfy
\begin{equation}\label{re:A0D0}
D_{0}(z)=-A_{0}(\zeta_1 z),
\end{equation}
\begin{equation}\label{re:B0C0}
C_{0}(z)=-B_{0}(\zeta_1^{-1} z),
\end{equation}
\begin{equation}
A_0(z)A_0(w) \simeq\ -2\ln z+\ln \zeta_1,
\end{equation}
\begin{equation}
B_0(z)B_0(w) \simeq\ -2\ln z-\ln \zeta_1,
\end{equation}
where  $A_0(z)A_0(w) \simeq\ -2\ln z+\ln \zeta_1$ means 
\begin{equation}\label{re:B0}
:\exp(A_0(z)): :\exp(A_0(w)):=\zeta_1 z^{-2} :\exp(A_0(z)+A_0(w)):
\end{equation} 
ect.. 

In principle, we can realize these relations with enough free fields. Our construction gives a realization of these relations. The irreducibility of this new free field representation remains an open question.
\subsection{Representations with level $(\zeta_1,\zeta_2)$}
For any $\zeta_1\neq 0$ and $\zeta_2\neq 0$ , we obtain representations of $\mathcal{U}$ with level $(\zeta_1,\zeta_2)$ by introducing modification factors:
\begin{equation}
E(z) \mapsto \frac{1}{z}[e_A(z):\exp(A(z)): +e_B(z):\exp(B(z)):],
\end{equation}
\begin{equation}
F(z) \mapsto \frac{1}{z}[ f_C(z):\exp(C(z)):+ f_D(z):\exp(D(z)):],
\end{equation}
\begin{equation}
K^{+}(z) \mapsto k^{+}(z) :\exp(A_{+}(\zeta_1^{1/2}z)+D_{+}(\zeta_1^{-1/2}z)):,
\end{equation}
\begin{equation}
K^{-}(z) \mapsto k^{-}(z) :\exp(B_{-}(\zeta_1^{-1/2}z)+C_{-}(\zeta_1^{1/2}z)):,
\end{equation}
where the modification factors $e(z)$, $f(z)$, $k^{\pm}(z)$ satisfy the following relations:
\begin{equation}
k^{+}(z)=e_A(\zeta_1^{1/2}z) f_D(\zeta_1^{-1/2}z) ,
\end{equation}  
\begin{equation}
k^{-}(z)=e_B(\zeta_1^{-/2}z) f_C(\zeta_1^{1/2}z) ,
\end{equation}  
due to relation (\ref{re:EF}) and the constant parts of $k^{\pm}(z)$ are fixed by $\zeta_2$. Assuming $C=\gamma^k$, we can obtain representations with level $(\gamma^k,\gamma^l)$ by choosing 
\begin{equation}
k^{\pm}(z)=\gamma^{\mp l} ,
\end{equation}
\begin{equation}
e_A(z)=e_B(z)=z^{-l/k} ,\quad f_C(z)=f_D(z)=z^{l/k},
\end{equation}
analogous to those in Section 2. Or we can choose all of them as constants:
\begin{equation}
k^{\pm}(z)=\gamma^{\mp l} ,
\end{equation}
\begin{equation}
e_A(z)=\gamma^{-l/2} ,\quad e_B(z)=\gamma^{l/2},
\end{equation}
\begin{equation}
f_C(z)=\gamma^{l/2} ,\quad f_D(z)=\gamma^{-l/2}.
\end{equation}
\subsection{Reconstruction of the previous free field realization}
In this subsection, we show that the horizontal representation presented in \cite{FHHSY10} can be recovered from our construction. For simplicity, let us fix the level to be $(\gamma,~1)$.  
Define 
\begin{align}
\tilde{\eta}(z)=
& \exp \left(\sum_{n > 0} \frac{1}{1-\gamma^2}(a^1_{-n}+\gamma^n a^2_{-n}+b^2_{-n}+\gamma^n c^1_{-n})z^n\right) 
\\ \notag
&\times \exp \left(\sum_{n > 0} \frac{1}{1-\gamma^2}(a^1_{n}+\gamma^n a^2_{n}+b^1_{n}+\gamma^n c^2_{n})z^{-n}\right),
\end{align}
and 
\begin{align}
\tilde{\xi}(z) =
& \exp \left(-\sum_{n > 0} \frac{\gamma^n}{1-\gamma^2}(a^1_{-n}+\gamma^n a^2_{-n}+b^2_{-n}+\gamma^n c^1_{-n})z^n\right) 
\\ \notag
&\times \exp \left(-\sum_{n > 0} \frac{\gamma^n}{1-\gamma^2}(a^1_{n}+\gamma^n a^2_{n}+b^1_{n}+\gamma^n c^2_{n})z^{-n}\right).
\end{align}
Then these two currents together with $\varphi^{+}(z)$ and $\varphi^{-}(z)$ give a free field realization of $\mathcal{U}'$ with level $(\gamma,1)$ under the following map:
\begin{equation}
E(z) \mapsto \tilde{\eta}(z),\quad 
F(z) \mapsto \tilde{\xi}(z) ,\quad
K^{+}(z) \mapsto \varphi^{+}(z),\quad
K^{-}(z) \mapsto \varphi^{-}(z),
\end{equation}
which is isomorphic to the horizontal representation given in Section 2.3. Although both $\tilde{\eta}(z)$ and $\eta(z)$ satisfy the same commuting relation (\ref{re:EE}):
\begin{align}
&(z-q_1w)(z-q_2w)(z-q_3w)\tilde{\eta}(z)\tilde{\eta}(w)\\
&=(z-q^{-1}_1w)(z-q^{-1}_2w)(z-q^{-1}_3w)\tilde{\eta}(w)\tilde{\eta}(z),
\end{align}
and
\begin{align}
&(z-q_1w)(z-q_2w)(z-q_3w)\eta(z)\eta(w)\\
&=(z-q^{-1}_1w)(z-q^{-1}_2w)(z-q^{-1}_3w)\eta(w)\eta(z),
\end{align}
they are quite different when calculating OPE relations.
The OPE of $\tilde{\eta}(z)$ and $\tilde{\eta}(w)$ is 
\begin{equation}
\tilde{\eta}(z)\tilde{\eta}(w)=S_3(w/z)^{-1} :\tilde{\eta}(z) \tilde{\eta}(w):.
\end{equation}
In the sense of analytic continuation, the structure function $g(w/z)$ is factorized as 
\begin{equation}
g(w/z)=S_3(w/z) S_3(z/w)^{-1}=S_3(w/z) S_3(q_3 w/z)^{-1},
\end{equation}
in this case.
However, $\eta(z)$ is decomposed into two terms and the OPE of $\eta(z)\eta(w)$ is 
\begin{align}
\eta(z)\eta(w)
=& \gamma \frac{z}{w} g_{+}(w/z)^{-1}  :\exp(A(z)+A(w)):
\\ \notag 
&+\frac{1}{wz}g(w/z)^{-1} :\exp(A(z)+B(w)): \\ \notag
&+ \frac{1}{wz}:\exp(B(z)+A(w)):
\\ \notag 
&+\gamma^{-1} \frac{z}{w} g_{+}(w/z)^{-1} :\exp(B(z)+B(w)):,
\end{align}
where different OPE factors appear, and the structure function is factorized through different factorizations. This leads to a difference in their singularity structures:
\begin{equation}
	\eta(w)\eta(z)-g(z/w)^{-1}\eta(z)\eta(w)=\sum_{i=1}^{3} \delta(q_i^{-1}w/z) \Lambda_i(w),
\end{equation}
while
\begin{equation}
\tilde{\eta}(w)\tilde{\eta}(z)-g(z/w)^{-1}\tilde{\eta}(z)\tilde{\eta}(w)=\sum_{j=1}^{2} \delta(q_j^{-1}w/z) \tilde{\Lambda}_j(w),
\end{equation}
where $\Lambda_i(w)\ (i=1,2,3)$ and $\tilde{\Lambda}_j(w)\ (j=1,2)$ are some fields. 

 We can regard that the two terms of $\eta(z)$ are obtained by twisting $\tilde{\eta}(z)$ in different directions. More precisely, we have
\begin{equation}
	:\exp(A(z)):= \alpha_A(z) \tilde{\eta}(z) \beta_A(z),
\end{equation}
and 
\begin{equation}
    :\exp(B(z)):= \alpha_B(z) \tilde{\eta}(z) \beta_B(z),
\end{equation}
for some half currents $\alpha_X(z)$ and $\beta_X(z)$ $(X=A,B)$ satisfying 
\begin{equation}
	[K^{\pm}(z),\alpha_X(w)]=[K^{\pm}(z),\beta_X(w)]=0.
\end{equation}
\par
We regard the factorization of the structure function $g(z)$ as the most crucial difference between our free field realization with the previous free field realization. This difference also appears in other defining relations.
\par
Note that oscillators $b^1_{n},c^2_{n},b^2_{-n},c^1_{-n}$ have no effects in the computations of OPE relations in the above reconstruction. That is, we have extra degrees of freedom in realizing the reconstruction. Another choice is given as follows: 
\begin{align}
E(z) \mapsto &\exp \left(\sum_{n > 0} \frac{1}{1-\gamma^2}(a^1_{-n}+\gamma^n a^2_{-n})z^n\right)
\\ \notag
&\times\exp \left(\sum_{n > 0} \frac{1}{1-\gamma^2}(a^1_{n}+\gamma^n a^2_{n})z^{-n}\right), 
\end{align}
\begin{align}
F(z) \mapsto &\exp \left(-\sum_{n > 0} \frac{\gamma^n}{1-\gamma^2}(a^1_{-n}+\gamma^n a^2_{-n})z^n\right)
\\ \notag
&\times\exp \left(-\sum_{n > 0} \frac{\gamma^n}{1-\gamma^2}(a^1_{n}+\gamma^n a^2_{n})z^{-n}\right), 
\end{align}
\begin{equation}
K^{+}(z) \mapsto \exp (\sum_{n > 0} (\gamma^{-n/2}a^1_{n}+\gamma^{n/2} a^2_{n})z^{-n}) ,
\end{equation}
\begin{equation}
K^{-}(z) \mapsto \exp (\sum_{n > 0} (\gamma^{-n/2}a^1_{-n}+\gamma^{n/2} a^2_{-n})z^{n}) .
\end{equation}

\subsection{Generalized Serre Relations}
We explain the meaning of the Serre relations within the framework of the free field construction. It demonstrates that our construction satisfies a generalized form (\ref{re:gSerreE}) of the Serre relations.

In the sense of analytic continuation, we can rewrite the commutation relation of $E(z)$ as 
\begin{equation}
E(z)E(w)=g(z,w)^{-1} E(w)E(z),
\end{equation}
where we denote 
\begin{equation*}
g(z,w)=\frac{(z-q_1w)(z-q_2w)(z-q_3w)}{(z-q^{-1}_1w)(z-q^{-1}_2w)(z-q^{-1}_3w)},
\end{equation*}
as a rational function. It can be formulated more precisely in the framework of free field construction. For instance,
\begin{align*}
(:&\exp(A(w)): +:\exp(B(w)):)(:\exp(A(z)): +:\exp(B(z)):)
\\
=&
g(z/w)^{-1}(:\exp(A(z)): +:\exp(B(z)):)(:\exp(A(w)): +:\exp(B(w)):)
\\
&+\gamma_1 \Delta_{+}(z/w):\exp(A(z)+A(w)): +\gamma_1^{-1} \Delta_{+}(z/w):\exp(B(z)+B(w)):
\\
&+\Delta_g(z/w) :\exp(A(z)+B(w)):,
\end{align*}
where
\begin{align}
\Delta_{g}(z/w) :=
&g(z/w)-g(w/z)^{-1}
\\ \notag
=
&\frac{(1-q_1^{-2})(1-q_2)(1-q_3)}{(1-q_1^{-1}q_2)(1-q_1^{-1}q_3)}\delta(q_1z/w)
\\ \notag
&+\frac{(1-q_2^{-2})(1-q_1)(1-q_3)}{(1-q_2^{-1}q_1)(1-q_2^{-1}q_3)}\delta(q_2z/w)
\\ \notag
&+\frac{(1-q_3^{-2})(1-q_1)(1-q_2)}{(1-q_3^{-1}q_1)(1-q_3^{-1}q_1)}\delta(q_3z/w),
\end{align}
and 
\begin{align}
\Delta_{+}(z/w) :=
&w^{-2}g_{+}(z/w)^{-1}-z^{-2}g_{-}(w/z)^{-1}
\\ \notag
=
&\frac{1-q_1^{-1}}{(1-q_1^{-1}q_2)(1-q_1^{-1}q_3)z^2}\delta(q_1z/w)
\\ \notag
&+\frac{1-q_2^{-1}}{(1-q_2^{-1}q_1)(1-q_2^{-1}q_3)z^2}\delta(q_2z/w)
\\ \notag
&+\frac{1-q_3^{-1}}{(1-q_3^{-1}q_1)(1-q_3^{-1}q_1)z^2}\delta(q_3z/w).
\end{align}
That is 
\begin{equation*}
E(w)E(z)=g(z/w)^{-1}E(z)E(w)+\sum_{i=1}^{3} \delta(q_i^{-1}w/z)\Lambda_i(w),
\end{equation*}
where $\Lambda_i(w)$ are some fields in the normal order. Thus we can use this to exchange all the terms in the Serre relations 
likes
\begin{equation*}
E(z_2)E(z_3)E(z_1),
\end{equation*}
into 
\begin{equation*}
E(z_1)E(z_2)E(z_3).
\end{equation*}
Then 
\begin{align*}
Sym&_{z_1,z_2,z_3} \frac{z_2}{z_3}[E(z_1),[E(z_2),E(z_3)]]
\\
=
& [\left(\frac{z_2}{z_3}+\frac{z_2}{z_1}-\frac{z_3}{z_2}-\frac{z_1}{z_2}\right)(1+g(z_1/z_2)^{-1}g(z_2/z_3)^{-1}g(z_1/z_3)^{-1})
\\ 
&+\left(\frac{z_3}{z_1}+\frac{z_3}{z_2}-\frac{z_1}{z_3}-\frac{z_2}{z_3}\right)
(g(z_1/z_2)^{-1}+g(z_1/z_3)^{-1}g(z_2/z_3)^{-1})
\\ 
&+ \left(\frac{z_1}{z_2}+\frac{z_1}{z_3}-\frac{z_2}{z_1}-\frac{z_3}{z_1}\right)
(g(z_2/z_3)^{-1}+g(z_1/z_2)^{-1}g(z_1/z_3)^{-1})]
\\ 
&~~~\times E(z_1)E(z_2)E(z_3)+\dots
\end{align*}
where $\dots$ stands for terms containing delta functions. The factor before $E(z_1)E(z_2)E(z_3)$ can be considered as the Laurent expansion of a rational function in the region $|z_3|\gg|z_2|\gg|z_1|$, which is equal to zero. Therefore
\begin{equation*}
Sym_{z_1,z_2,z_3} \frac{z_2}{z_3}[E(z_1),[E(z_2),E(z_3)]] 
=\sum_{i,j,k}\delta(q_iz_j/z_k)\Lambda_i^{j,k}(z_1,z_2,z_3) 
\end{equation*}
where $\Lambda_i^{j,k}(z_1,z_2,z_3) $ contains operators. Note that there might many ways to split it into such a sum of delta functions.  
Nevertheless, it shows that there exist polynomials $f(z_1,z_2,z_3)\in\mathbb{C}[z_1,z_2,z_3]$ such that
\begin{equation}\label{re:gSerreE}
f(z_1,z_2,z_3)Sym_{z_1,z_2,z_3} \frac{z_2}{z_3}[E(z_1),[E(z_2),E(z_3)]]=0.
\end{equation}
One of them is the following polynomials of degree-9
\begin{equation*}
f(z_1,z_2,z_3)=\prod_{i=1}^{3} (z_1-q_i^{-1}z_2)(z_2-q_i^{-1}z_3)(z_1-q_i^{-1}z_3),
\end{equation*}
which is derived from the above discussion:
\begin{align*}
Sym&_{z_1,z_2,z_3} \frac{z_2}{z_3}[E(z_1),[E(z_2),E(z_3)]]
\\
=
&\sum_{i=1}^3\delta(q_iz_1/z_2)\Lambda_i^{1,2}(z_1,z_2,z_3) +\sum_{i=1}^3\delta(q_iz_2/z_3)\Lambda_i^{2,3}(z_1,z_2,z_3)
\\
&+\sum_{i=1}^3\delta(q_iz_1/z_3)\Lambda_i^{1,3}(z_1,z_2,z_3).
\end{align*}
Since the $z_i's$ are symmetric under the action of the permutation group $S_3$, there are six relations that can be derived directly from the commutation relations of $E(z)$. There may be other $f(z_1,z_2,z_3)\in\mathbb{C}[z_1,z_2,z_3]$, which depends on the specific realization. 
All such polynomials $f(z_1,z_2,z_3)$ form an ideal $I\subset \mathbb{C}[z_1,z_2,z_3]$. Each $f(z_1,z_2,z_3)\in I$ corresponds to a cubic relation of the form (\ref{re:gSerreE}). Since $\mathbb{C}[z_1,z_2,z_3]$ is a Noetherian ring, the ideal $I$ is finitely generated. Consequently, there are only finitely many such independent relations. From this perspective, we regard this condition
\begin{equation}
f(z_1,z_2,z_3)Sym_{z_1,z_2,z_3} \frac{z_2}{z_3}[E(z_1),[E(z_2),E(z_3)]]=0
,\quad f(z_1,z_2,z_3)\in I
\end{equation}
as a generalized form of the Serre relations. Specifically, the Serre relation (\ref{re:Serre_E}) enforces the cancellation of terms containing formal delta functions. In other words, 
\begin{equation*}
I=\mathbb{C}[z_1,z_2,z_3].
\end{equation*}

After substituting (\ref{def:eta}) into the left hand side of (\ref{re:Serre_E}), we find that there are a total of 192 terms. These 192 terms are classified into eight sets, with each set containing 24 terms. Terms with the same normal product after normal ordering are collected together, such as:
\begin{align} \label{eq:BAA}
\left(\frac{z_2}{z_3}+\frac{z_2}{z_1}-\frac{z_3}{z_2}-\frac{z_1}{z_2}\right) 
&(:\exp(B(z_1))::\exp(A(z_2))::\exp(A(z_3)):
\\ \notag
& +:\exp(A(z_3))::\exp(A(z_2))::\exp(B(z_1)):)
\\ \notag
~~~~~~~~~~+\left(\frac{z_1}{z_2}+\frac{z_1}{z_3}-\frac{z_2}{z_1}-\frac{z_3}{z_1}\right) 
&(:\exp(A(z_2))::\exp(B(z_1))::\exp(A(z_3)):
\\ \notag
& +:\exp(A(z_3))::\exp(B(z_1))::\exp(A(z_2)):)
\\ \notag
~~~~~~~~~~+\left(\frac{z_3}{z_1}+\frac{z_3}{z_2}-\frac{z_1}{z_3}-\frac{z_2}{z_3}\right)
&(:\exp(B(z_1))::\exp(A(z_1))::\exp(A(z_3)):
\\ \notag
& +:\exp(A(z_2))::\exp(A(z_3))::\exp(B(z_1)):),		    
\end{align}
where we pull out a common factor $ z_1^{-1}z_2^{-1}z_3^{-1}$.
All these terms have the same normal product
\begin{equation*}
:\exp(B(z_1)+A(z_2)+A(z_3)):.
\end{equation*}
After extracting a factor $\gamma_1$, the OPE factor, denoted by $h_{BAA}(z_1,z_2,z_3)$, is 
\begin{align}
h&_{BAA}(z_1,z_2,z_3)
\\ \notag
=&\left(\frac{z_2}{z_3}+\frac{z_2}{z_1}-\frac{z_3}{z_2}-\frac{z_1}{z_2}\right) 
(z_2^{-2}g_{+}(z_3/z_2)^{-1}+z_3^{-2}g_{+}(z_2/z_3)^{-1}g(z_1/z_3)^{-1}g(z_1/z_2)^{-1})
\\ \notag
&+\left(\frac{z_1}{z_2}+\frac{z_1}{z_3}-\frac{z_2}{z_1}-\frac{z_3}{z_1}\right) 
(z_2^{-2}g_{+}(z_3/z_2)^{-1}g(z_1/z_2)^{-1}+z_3^{-2}g_{+}(z_2/z_3)^{-1}g(z_1/z_3)^{-1})
\\ \notag
&+\left(\frac{z_3}{z_1}+\frac{z_3}{z_2}-\frac{z_1}{z_3}-\frac{z_2}{z_3}\right)
(z_3^{-2}g_{+}(z_2/z_3)^{-1}+z_2^{-2}g_{+}(z_3/z_2)^{-1}g(z_1/z_2)^{-1}g(z_1/z_3)^{-1}).
\end{align}
The other terms are handled in the same way and we denote the OPE factor corresponding to 
\begin{equation*}
:\exp(X(z_1)+Y(z_2)+Z(z_3)):
\end{equation*}
by $h_{XYZ}(z_1,z_2,z_3)$ where
\begin{equation*}
XYZ\in\{ AAA,AAB,ABA,BAA,ABB,BAB,BBA,BBB  \}.
\end{equation*}
Then 
\begin{align}
h&_{AAA}(z_1,z_2,z_3)=h_{BBB}(z_1,z_2,z_3)
\\ \notag
=&\left(\frac{z_2}{z_3}+\frac{z_2}{z_1}-\frac{z_3}{z_2}-\frac{z_1}{z_2}\right) 
(z_1^{-4}z_2^{-2}g_{+}(z_3/z_2)^{-1}g_{+}(z_2/z_1)^{-1}g_{+}(z_3/z_1)^{-1}
\\ \notag
&+z_3^{-4}z_2^{-2}g_{+}(z_1/z_2)^{-1}g_{+}(z_3/z_1)^{-1}g_{+}(z_3/z_2)^{-1})
\\ \notag
&+\left(\frac{z_1}{z_2}+\frac{z_1}{z_3}-\frac{z_2}{z_1}-\frac{z_3}{z_1}\right) 
(z_2^{-4}z_1^{-2}g_{+}(z_1/z_2)^{-1}g_{+}(z_3/z_1)^{-1}g_{+}(z_3/z_2)^{-1}
\\ \notag
&+z_3^{-4}z_1^{-2}g_{+}(z_1/z_3)^{-1}g_{+}(z_2/z_1)^{-1}g_{+}(z_2/z_3)^{-1})
\\ \notag
&+\left(\frac{z_3}{z_1}+\frac{z_3}{z_2}-\frac{z_1}{z_3}-\frac{z_2}{z_3}\right)
(z_1^{-4}z_3^{-2}g_{+}(z_3/z_1)^{-1}g_{+}(z_2/z_3)^{-1}g_{+}(z_2/z_1)^{-1}
\\ \notag
&+z_2^{-4}z_3^{-2}g_{+}(z_3/z_2)^{-1}g_{+}(z_1/z_3)^{-1}g_{+}(z_1/z_2)^{-1}).
\end{align}
As for the crossing terms, $h_{ABA}(z_1,z_2,z_3)$ and $h_{AAB}(z_1,z_2,z_3)$ can be obtained by permuting $z_i's$ in $h_{BAA}(z_1,z_2,z_3)$:
\begin{equation}
h_{ABA}(z_1,z_2,z_3)=h_{BAA}(z_2,z_1,z_3),
\end{equation}
\begin{equation}
h_{AAB}(z_1,z_2,z_3)=h_{BAA}(z_3,z_2,z_1).
\end{equation}
Similarly, $h_{BAB}(z_1,z_2,z_3)$ and $h_{BBA}(z_1,z_2,z_3)$ can be expressed in terms of $h_{ABB}(z_1,z_2,z_3)$:
\begin{equation}
h_{BAB}(z_1,z_2,z_3)=h_{ABB}(z_2,z_1,z_3),
\end{equation}
\begin{equation}
h_{BBA}(z_1,z_2,z_3)=h_{ABB}(z_3,z_2,z_1),
\end{equation}
where
\begin{align*}
h&_{ABB}(z_1,z_2,z_3)
\\
=&\left(\frac{z_2}{z_3}+\frac{z_2}{z_1}-\frac{z_3}{z_2}-\frac{z_1}{z_2}\right) 
(z_2^{-2}g_{+}(z_3/z_2)^{-1}g(z_2/z_1)^{-1}g(z_3/z_1)^{-1}+z_3^{-2}g_{+}(z_2/z_3)^{-1})
\\
&+\left(\frac{z_1}{z_2}+\frac{z_1}{z_3}-\frac{z_2}{z_1}-\frac{z_3}{z_1}\right) 
(z_2^{-2}g_{+}(z_3/z_2)^{-1}g(z_3/z_1)^{-1}+z_3^{-2}g_{+}(z_2/z_3)^{-1}g(z_2/z_1)^{-1})
\\
&+\left(\frac{z_3}{z_1}+\frac{z_3}{z_2}-\frac{z_1}{z_3}-\frac{z_2}{z_3}\right)
(z_3^{-2}g_{+}(z_2/z_3)^{-1}g(z_2/z_1)^{-1}g(z_3/z_1)^{-1}+z_2^{-2}g_{+}(z_3/z_2)^{-1}).
\end{align*}
The symmetry between $h_{BAA}(z_1,z_2,z_3)$ and $h_{ABB}(z_1,z_2,z_3)$ is equivalent to 
\begin{equation*}
q_i \leftrightarrow q_i^{-1} ,~\quad i=1,2,3.
\end{equation*} 
After intricate calculations, of which details are presented in the appendix, we get
\begin{equation*}
h_{BAA}(z_1,z_2,z_3)=0,	
\end{equation*}
while
\begin{align} \label{eq:hAAA}
h&_{AAA}(z_1,z_2,z_3)
\\ \notag
=
&(q_1-q_2)^{-1}(q_1-q_3)^{-1}(q_2-q_3)^{-1}z_1^{-2}z_2^{-2}z_3^{-2}
\\ \notag
&\times
[\delta(q_1z_2/z_3)\delta(q_3^{-1}z_1/z_3)+\delta(q_2z_2/z_3)\delta(q_1^{-1}z_1/z_3)+\delta(q_3z_2/z_3)\delta(q_2^{-1}z_1/z_3)
\\ \notag
&+\delta(q_1z_1/z_3)\delta(q_3^{-1}z_2/z_3)+\delta(q_2z_1/z_3)\delta(q_1^{-1}z_2/z_3)+\delta(q_3z_1/z_3)\delta(q_2^{-1}z_2/z_3)
\\ \notag
&-\delta(q_1z_2/z_3)\delta(q_2^{-1}z_1/z_3)-\delta(q_2z_2/z_3)\delta(q_3^{-1}z_1/z_3)-\delta(q_3z_2/z_3)\delta(q_1^{-1}z_1/z_3)
\\ \notag
&-\delta(q_1z_1/z_3)\delta(q_2^{-1}z_2/z_3)-\delta(q_2z_1/z_3)\delta(q_3^{-1}z_2/z_3)-\delta(q_3z_1/z_3)\delta(q_1^{-1}z_2/z_3)
].
\end{align}
Thus
\begin{align}\label{eq:symE}
Sym&_{z_1,z_2,z_3} \frac{z_2}{z_3}[\eta(z_1),[\eta(z_2),\eta(z_3)]]
\\ \notag
=
&\frac{h_{AAA}(z_1,z_2,z_3)}{z_1z_2z_3}[\gamma_1^3:\exp(A(z_1)+A(z_2)+A(z_3)):
\\ \notag
&~~~~~~~~~~~~~~~~~~~~+\gamma_1^{-3}:\exp(B(z_1)+B(z_2)+B(z_3)):].
\end{align}

A similar discussion shows that
\begin{align}
Sym&_{z_1,z_2,z_3} \frac{z_2}{z_3}[\xi(z_1),[\xi(z_2),\xi(z_3)]]
\\ \notag
=
&-\frac{h_{AAA}(z_1,z_2,z_3)}{z_1z_2z_3}[\gamma_1^3:\exp(C(z_1)+C(z_2)+C(z_3)):
\\ \notag
&~~~~~~~~~~~~~~~~~~~~~~~~+\gamma_1^{-3}:\exp(D(z_1)+D(z_2)+D(z_3)):].
\end{align}

All the polynomials $f(z_1,z_2,z_3)\in\mathbb{C}[z_1,z_2,z_3]$ satisfy 
\begin{equation*}
f(z_1,z_2,z_3)h_{AAA}(z_1,z_2,z_3)=0,
\end{equation*}
constitute an ideal $I_h\subset \mathbb{C}[z_1,z_2,z_3]$. To determine this ideal, let us consider each term in $h_{AAA}(z_1,z_2,z_3)$ separately. The products of delta functions 
\begin{equation*}
\delta(q_iz_1/z_3)\delta(q_j^{-1}z_2/z_3) \quad i\neq j,
\end{equation*} 
are canceled by 
\begin{equation*}
z_3-q_iz_1,\ z_3-q_j^{-1}z_2 
\end{equation*}
which generates a prime ideal $I_{ij}\subset \mathbb{C}[z_1,z_2,z_3]$. Then $I_h$ is the intersection of all these prime ideals:
\begin{equation*}
I_h=\cap I_{ij}.
\end{equation*}
Equivalently, the polynomial $f\in I_h$ is characterized by the following property: for $\{i,j,k\}=\{1,2,3\}$, if
	\begin{equation}
	\left\{ \frac{x_i}{x_j},\frac{x_j}{x_k},\frac{x_k}{x_i}\right\}=\{q_1,q_2,q_3\},
	\end{equation}  
then
	\begin{equation*}
	f(x_1,x_2,x_3)=0,
	\end{equation*}
which is similar to the wheel condition in the shuffle algebra \cite{FT}.
Although $I_h$ is not a prime ideal, it is finitely generated. In principle, a set of generators can be computed using the Gröbner basis techniques, although the computation is quite involved. Therefore, we present several specific elements of this ideal, rather than providing a complete generating set. The lowest-degree polynomial among them is the following degree-3 symmetric function:
\begin{equation}
\begin{aligned}
(q_1&+q_2+q_3+q_1^{-1}+q_2^{-1}+q_3^{-1})m_{(1^3)}(z_1,z_2,z_3)-m_{(21)}(z_1,z_2,z_3)
\\
=&
(q_1+q_2+q_3+q_1^{-1}+q_2^{-1}+q_3^{-1})z_1z_2z_3
\\
&-(z_1^2z_2+z_1^2z_3+z_1z_2^2+z_2^2z_3+z_1z_3^2+z_2z_3^2).
\end{aligned}
\end{equation}
The corresponding cubic relation is equivalent to 
\begin{align*}
Sym_{z_1,z_2,z_3}& \left(\frac{z_2}{z_3}+\frac{z_2^2}{z_3^2}+\frac{z_2^2}{z_1z_3}+\frac{z_1z_2}{z_3^2}-(q_1+q_2+q_3+q_1^{-1}+q_2^{-1}+q_3^{-1})\frac{z_2}{z_3}\right)
\\
&\times[E(z_1),[E(z_2),E(z_3)]]=0,
\end{align*}
which follows from a direct computation using
\begin{align*}
Sym&_{z_1,z_2,z_3} f(z_1,z_2,z_3) [E(z_1),[E(z_2),E(z_3)]]
\\
=
&Sym_{z_1,z_2,z_3} (f(z_1,z_2,z_3)+f(z_3,z_2,z_1)-f(z_1,z_3,z_2)-f(z_3,z_1,z_2))
\\
&\times E(z_1)E(z_2)E(z_3).
\end{align*}
Note that the way to rewrite it in the form of 
\begin{equation*}
Sym_{z_1,z_2,z_3} f(z_1,z_2,z_3) [E(z_1),[E(z_2),E(z_3)]]=0,
\end{equation*}
is not unique since there are some $f(z_1,z_2,z_3)\neq 0$ such that
\begin{equation*}
f(z_1,z_2,z_3)+f(z_3,z_2,z_1)-f(z_1,z_3,z_2)-f(z_3,z_1,z_2)=0.
\end{equation*} 

Furthermore, this result can be compared with the Serre relations appearing in horizontal representations. 
From
\begin{equation*}
S_3(w/z)^{-1}=(z-q_3w)(z-q_3^{-1}w) z^{-2}g_{+}(w/z)^{-1},
\end{equation*}
and 
\begin{equation*}
Sym_{z_1,z_2,z_3} \left (\frac{z_2}{z_1}+\frac{z_2}{z_3}-\frac{z_1}{z_2}-\frac{z_3}{z_2} \right) S(z_2/z_1)S(z_3/z_1)S(z_3/z_2)=0,
\end{equation*}
we obtain
\begin{equation*}
\begin{aligned}
&(z_1-q_3z_2)(z_1-q_3^{-1}z_2)(z_1-q_3z_3)(z_1-q_3^{-1}z_3)(z_2-q_3z_3)(z_2-q_3^{-1}z_3)
\\
&\times Sym_{z_1,z_2,z_3}\left (\frac{z_2}{z_1}+\frac{z_2}{z_3}-\frac{z_1}{z_2}-\frac{z_3}{z_2}\right) z_1^{-4}z_2^{-2}g_{+}(z_3/z_2)^{-1}g_{+}(z_2/z_1)^{-1}g_{+}(z_3/z_1)^{-1}=0,
\end{aligned}
\end{equation*}
which agrees with the fact that 
\begin{equation*}
(z_1-q_3z_2)(z_1-q_3^{-1}z_2)(z_1-q_3z_3)(z_1-q_3^{-1}z_3)(z_2-q_3z_3)(z_2-q_3^{-1}z_3) \in I_h.
\end{equation*}

Note that $h_{AAA}(z_1,z_2,z_3)$ consists purely of $g_{+}(z)$ while $h_{BAA}(z_1,z_2,z_3)$ is a mixed expression containing both $g_{+}(z)$ and $g(z)$. These identities as Laurent series may relate to some combinatorial identities, such as those in \cite{Jing98}. We also hope that their meaning in terms of state counting will be clarified in the future, as with the Serre relations in the Yangian \cite{LY}.

\section{Intertwiner}
The intertwiner is an important algebraic object in Conformal Field Theory (CFT), it can be related to gauge theories in the context of AGT duality. The vector intertwiner, which acts on a tensor product of the vertical vector representation and the free field horizontal representation, was first introduced \cite{FHHSY10}. It was shown that the vector intertwiner can reproduce the holomorphic blocks of 3d quiver gauge theories  \cite{Ze}. Also, it can be seen \cite{AKMMSZ} for a uniform approach to constructing various intertwiners of the DIM algebra by using the vector intertwiners. To generalize these intertwiners to a more general setting, we replace the free field representations of Feigin-Hashizume-Hoshino-Shiraishi-Yanagida by our free field representations. With this change, the first level of $\mathcal{U}$ can be an arbitrary value $\zeta_1\in\mathbb{C}^{\times}$ instead of a special quantized value $\gamma$ in the construction of intertwiners. As the first step in this process, we fix the vertical representation to be the vector representation $V_1(u)$. There are two types of intertwiners, which play the role of creation operators and annihilation operators, respectively. One type is called the intertwiner, then the other type is called the dual intertwiner.
\subsection{Vector intertwiner} 
In this subsection, we consider the trivalent intertwiner $\Phi(u):V_1(u)\otimes V \rightarrow V'$ , which is defined by the intertwining condition 
\begin{equation}
a\Phi(u)=\Phi(u) \Delta(a) , \qquad  \forall a \in \mathcal{U}.
\end{equation} 
Recall that $V_1(u)$ has a basis $\{[u]_i  | i\in\mathbb{Z}\}$ which simultaneously diagonalizes $K^{\pm}(z)$. We can introduce the $n$-component $\Phi_{n}(u)$ of this intertwiner such that $\Phi_{n} : V \rightarrow V'$ is considered as an operator acting on $V$ and 
\begin{align}
\Phi_{n}(u) (v) = \Phi(u)([u]_n \otimes v), \qquad \forall v\in V.
\end{align}
Applying $C_1=1$ and $C_2=\zeta_1$ to the defining relations of the coproduct (\ref{re:coE})-(\ref{re:coKp}), the intertwining relations for the $n$-component can be read as 
\begin{equation}
\langle n+1 | E(z) | n \rangle \Phi_{n+1}(u) + \langle n | K^{-}(z) | n \rangle \Phi_{n}(u) \rho (E(z)) = \rho'(E(z)) \Phi_{n}(u),
\end{equation}
\begin{equation}
\langle n-1 | F(z) | n \rangle \Phi_{n-1}(u) \rho (K^{+}(\zeta_1^{1/2}z)) + \Phi_{n}(u) \rho (F(z)) = \rho'(F(z)) \Phi_{n}(u),
\end{equation} 
\begin{equation}
\langle n | K^{+}(\zeta_1^{1/2} z) | n \rangle \Phi_{n}(u) \rho(K^{+}(z)) = \rho'(K^{+}(z)) \Phi_{n}(u) ,
\end{equation}
\begin{equation}
\langle n | K^{-}(\zeta_1^{-1/2} z) | n \rangle \Phi_{n}(u) \rho(K^{-}(z)) = \rho'(K^{-}(z)) \Phi_{n}(u) .
\end{equation}
Here $\langle m | X(z) | n \rangle $ denotes the matrix elements defined by: 
\begin{equation}
X(z) [u]_n = \sum_{m \in \mathbb{Z}} \langle m | X(z) | n \rangle [u]_m
\end{equation} 
in the vector representation $V_1(u)$.
$\rho: \mathcal{U}\mapsto End(V)$ and $\rho: \mathcal{U}\mapsto End(V')$ stand for homomorphisms in vertex operator representations $(V,\rho)$ and $(V',\rho')$ respectively. More precisely, these relations can be written as 
\begin{align} \label{re:int1E}
& \frac{1}{z} (e_A'(z):\exp(A(z)): +e_B'(z):\exp(B(z)):) \Phi_{n}(u) 
\\ \notag
&-  \frac{1}{z}  S_1(q^{-n} z/u) \Phi_{n}(u) (e_A(z):\exp(A(z)): +e_B(z):\exp(B(z)):) 
\\ \notag
&=  \mathcal{E} \delta(q_1^{n+1}u/z) \Phi_{n+1}(u),
\end{align}
\begin{align} \label{re:int1F}
& \frac{1}{z} (f_C'(z):\exp(C(z)):+ f_D'(z):\exp(D(z)):) \Phi_{n}(u)
\\ \notag
&- \frac{1}{z} \Phi_{n}(u) (f_C(z):\exp(C(z)):+ f_D(z):\exp(D(z)):)
\\ \notag
&=  \mathcal{F} \delta(q_1^{n}u/z) \Phi_{n-1}(u) k^{+} (\zeta_1^{1/2} z) \varphi^{+} (\zeta_1^{1/2} z),
\end{align}
\begin{equation} \label{re:int1Kp}
S_1(\zeta_1^{-1/2} q_1^{n+1} u/z) k^{+}(z) \Phi_{n}(u) \varphi^{+}(z) = k^{+,\prime}(z) \varphi^{+}(z) \Phi_{n}(u),
\end{equation}
\begin{equation} \label{re:int1Kn}
S_1(\zeta_1^{-1/2} q_1^{-n} z/u) k^{-}(z) \Phi_{n}(u) \varphi^{-}(z) = k^{-,\prime}(z) \varphi^{-}(z) \Phi_{n}(u).
\end{equation}
Observing relation (\ref{re:int1E}) carefully, we will find that it can be considered as a recursion relation to derive $\Phi_{n+1}(u)$ from $\Phi_{n}(u)$. Similarly, relation (\ref{re:int1F}) can be considered as a recursion relation to derive $\Phi_{n-1}(u)$ from $\Phi_{n}(u)$. 
\par
To state our construction, we need to introduce the theta function. For $|q|<1$, define 
\begin{equation}
(z;q)_{\infty}=\prod_{n=0}^{\infty} (1-q^n z)=\exp\left(-\sum_{r=1}^{\infty} \frac{z^r}{r(1-q^r)}\right),
\end{equation}
and the theta function is defined as 
\begin{equation}
\theta_{q}(z)=(z;q)_{\infty}(qz^{-1};q)_{\infty}.
\end{equation} 
For $|q|>1$, the infinite product $(z;q)_{\infty}$ can be defined via the analytic continuation:
\begin{equation}
(z;q)_{\infty}=(q^{-1}z;q)_{\infty}^{-1}.
\end{equation}
And the theta function is given by
\begin{equation}
\theta_{q}(z)=\theta_{q^{-1}}(q^{-1}z)^{-1}.
\end{equation}
The theta function satisfies
\begin{equation}\label{prop:theta}
\theta_{q}(q^nz)=(-z)^{-n}q^{-\frac{n(n-1)}{2}}\theta_{q}(z).
\end{equation}
For convenience, we give the construction of $\Phi(u)$ for $|q_1|<1$ and the statement for $|q_1|>1$ can be obtained through analytic continuation.
\subsubsection{Constructions of $\Phi_{n}(u)$}
We aim to construct the $n$-component $\Phi_{n}(u)$ of the intertwiner and specify the relative conditions between factors such as $e_A(z)$ and $e_A'(z)$ in the vertex operator representations. We use the vertex operator $:\exp(A(z)):$ as the main building block in the construction of $\Phi_{n}(u)$. Define: 
\begin{align}
\tilde{\Phi}_{n}(u)= 
& \exp\left(-\sum_{r=1} \frac{q_1^{r(n+1)}}{1-q_1^{r}}((b^1_{-r}-\zeta_1^{-r}c^2_{-r})z^r)\right)
\\ \notag
& \times \exp\left((p^2_c-p^1_b)\left(\frac{(n+1)n}{2}\ln q_1 +n\ln u-\frac{n}{2}\ln\zeta_1\right)+n(q^1_b-q^2_c)\right)
\\ \notag 
& \times \exp\left(\sum_{r=1} \frac{q_1^{-rn}}{1-q_1^{r}}(a^1_{r}+b^1_{r})z^{-r}\right),
\end{align}
and specially for $n=0$, we have
\begin{align}
\tilde{\Phi}_0(u)=\exp\left(-\sum_{r=1}\frac{q_1^r}{1-q_1^r}(b^1_{-r}-\zeta_1^{-r}c^2_{-r})u^r\right) \exp\left(\sum_{r=1}\frac{1}{1-q_1^r}(a_r^1+b_r^1)u^{-r}\right).
\end{align}
Note that $\tilde{\Phi}_{n}(u)$ is equal to $\tilde{\Phi}_0(q_1^n u)$ up to the zero mode part and satisfies the following recursion relation:
\begin{equation} \label{re:recursionA}
\tilde{\Phi}_{n+1}(u) = :\tilde{\Phi}_n (u) \exp(A(q_1^{n+1} u)):  .
\end{equation}
Formally, we can express $\tilde{\Phi}_{n}(u)$ as an infinite product: 
\begin{equation}
: \prod_{j=n}^{\infty} \exp(-A_{-}(q_1^{j}u)) \exp(-A_{+}(q_1^{j}u)) \prod_{j=1}^{n} \exp(A_{0}(q_1^{j}u)): ,
\quad n\geq 0
\end{equation}
and 
\begin{equation}
:\prod_{j=n}^{\infty} \exp(-A_{-}(q_1^{j}u)) \exp(-A_{+}(q_1^{j}u)) \prod_{j=n}^{-1} \exp(-A_{0}(q_1^{j+1}u)): ,
\quad n \leq 0.
\end{equation}
We want to construct the $n$-component $\Phi_{n}(u)$ of form:
\begin{equation}
\Phi_{n}(u)=t_n(u) \tilde{\Phi}_{n}(u),
\end{equation}
where $t_n(u)$ is a factor that will be determined from relations (\ref{re:int1E}) or (\ref{re:int1F}) recursively. 
\par
Before discussing general cases, let us start by computing the OPE relations between $\tilde{\Phi}_{n}(u)$ and $:\exp(A(z)):,:\exp(B(z)):,:\exp(C(z)):,:\exp(D(z)):$ in the case of $n=0$, and explaining the roles that they play in the relations (\ref{re:int1E}) and (\ref{re:int1F}). Since $:\exp(A(z)):$ is the main building block of $\tilde{\Phi}_{n}(u)$ and $\tilde{\Phi}_0(u)$ has trivial zero mode part, these OPE relations can be derived from relations (\ref{re:AnDn})-(\ref{re:BpAn}). The OPE relations between $\tilde{\Phi}_{0}(u)$ and $:\exp(A(z)):$ are given by
\begin{equation}\label{ope:A_Phi_0}
:\exp(A(z)): \tilde{\Phi}_{0}(u) =\frac{(q_1q_2u/z;q_1)_{\infty}(q_1q_3u/z;q_1)_{\infty}}{1-q_1u/z}:\exp(A(z)) \tilde{\Phi}_{0}(u):,
\end{equation}
\begin{equation}
\tilde{\Phi}_{0}(u) :\exp(A(z)): = \frac{1-z/u}{(q_2z/u;q_1)_{\infty}(q_3z/u;q_1)_{\infty}} :\exp(A(z)) \tilde{\Phi}_{0}(u):.
\end{equation}
Then 
\begin{equation}\label{ope:S_Phi_0_A}
S_1(z/u)\tilde{\Phi}_{0}(u) :\exp(A(z)): = \frac{:\exp(A(z)) \tilde{\Phi}_{0}(u):}{(1-q_1^{-1}z/u)(q_1q_2z/u;q_1)_{\infty}(q_1q_3z/u;q_1)_{\infty}} .
\end{equation}
Comparing (\ref{ope:A_Phi_0}) and (\ref{ope:S_Phi_0_A}), observe that 
\begin{align}
&(q_1q_2u/z;q_1)_{\infty}(q_1q_3u/z;q_1)_{\infty} (q_1q_2z/u;q_1)_{\infty}(q_1q_3z/u;q_1)_{\infty}\\
&= \theta_{q_1}(q_1q_2u/z)\theta_{q_1}(q_1q_3u/z).
\end{align}
If we choose
\begin{equation}
e_A'(z)=(-q_1u/z) \theta_{q_1}(q_1q_2u/z)^{-1}\theta_{q_1}(q_1q_3u/z)^{-1}e_A(z),
\end{equation}
then terms containing $:\exp(A(z)):$ in (\ref{re:int1E}) are:
\begin{align} 
& \frac{1}{z} (e_A'(z):\exp(A(z)):  \tilde{\Phi}_{0}(u) - S_1(z/u) \tilde{\Phi}_{0}(u) e_A(z):\exp(A(z)):)  
\\ \notag
&= -\frac{e_A(z)}{z(q_1q_2z/u;q_1)_{\infty}(q_1q_3z/u;q_1)_{\infty}}  \delta(q_1u/z):\exp(A(z)) \tilde{\Phi}_{0}(u):
\\ \notag
& =  - \frac{e_A(q_1u)}{q_1u(q_1^2q_2;q_1)_{\infty}(q_1^2q_3;q_1)_{\infty}}\delta(q_1u/z) \tilde{\Phi}_{1}(u),
\end{align}
which illustrates the recursion relation (\ref{re:recursionA}).
Next, the OPE relations of $\tilde{\Phi}_{0}(u)$ and $:\exp(B(z)):$ are given by
\begin{equation}
\tilde{\Phi}_{0}(u) :\exp(B(z)): = S_1(z/u)^{-1} :\tilde{\Phi}_{0}(u) \exp(B(z)): ,
\end{equation}
\begin{equation}
:\exp(B(z)):\tilde{\Phi}_{0}(u) = :\tilde{\Phi}_{0}(u) \exp(B(z)): .
\end{equation}
If we set
\begin{equation}
e_B'(z)=e_B(z),
\end{equation}
these two terms cancel each other in (\ref{re:int1E}):
\begin{equation}
e_B'(z):\exp(B(z)):\tilde{\Phi}_{0}(u) -S_1(z/u)e_B(z)\tilde{\Phi}_{0}(u) :\exp(B(z)):=0.
\end{equation}
Similarly, 
\begin{equation}
:\exp(C(z)): \tilde{\Phi}_{0}(u) =:\exp(C(z)) \tilde{\Phi}_{0}(u): = \tilde{\Phi}_{0}(u) :\exp(C(z)):.
\end{equation}
so these two terms can be cancelled each other in (\ref{re:int1F}): 
\begin{equation}
f_C'(z):\exp(C(z)):\tilde{\Phi}_{0}(u) -f_C(z)\tilde{\Phi}_{0}(u) :\exp(C(z)):=0.
\end{equation}
with 
\begin{equation}
f_C'(z)=f_C(z).
\end{equation}
At last, the OPE relations between $\tilde{\Phi}_{0}(u)$ and $:\exp(D(z)):$ are given by
\begin{equation}
:\exp(D(z)): \tilde{\Phi}_{0}(u) =\frac{:\tilde{\Phi}_{0}(u)\exp(D(z)):}{(1-\zeta_1^{-1}u/z)(\zeta_1^{-1}q_1^2q_2u/z;q_1)_{\infty}(\zeta_1^{-1}q_1^2q_3u/z;q_1)_{\infty}} ,
\end{equation}
\begin{equation}
\tilde{\Phi}_{0}(u) :\exp(D(z)):= \frac{(\zeta_1q_2z/u;q_1)_{\infty}(\zeta_1q_3z/u;q_1)_{\infty}}{1-\zeta_1z/u} :\tilde{\Phi}_{0}(u)\exp(D(z)):.
\end{equation}
With a relative shift of factors 
\begin{equation}
f_D'(z)=(-\zeta_1^{-1}u/z)\theta_{q_1}(\zeta_1^{-1}q_1^2q_2u/z)\theta_{q_1}(\zeta_1^{-1}q_1^2q_3u/z)f_D(z),
\end{equation}
we have 
\begin{align} 
& \frac{1}{z} (f_D'(z):\exp(D(z)):  \tilde{\Phi}_{0}(u) - \tilde{\Phi}_{0}(u) f_D(z):\exp(D(z)):)  
\\ \notag
& = -\frac{1}{z}(\zeta_1q_2z/u;q_1)_{\infty}(\zeta_1q_3z/u;q_1)_{\infty}f_D(z)  \delta(\zeta_1^{-1}u/z) :\exp(D(z)) \tilde{\Phi}_{0}(u):
\\ \notag
&= - \zeta_1 u^{-1}e_A(u)^{-1}(q_2;q_1)_{\infty}(q_3;q_1)_{\infty}\delta(\zeta_1^{-1}u/z) \tilde{\Phi}_{-1}(u) k^{+}(\zeta_1^{-1/2}u)\varphi^{+}(\zeta_1^{-1/2}u) .
\end{align}
The last equality holds due to 
\begin{equation}
\varphi^{+}(\zeta_1^{-1/2}u)=:\exp(A(u))\exp(D(\zeta_1^{-1}u)):,
\end{equation}
and
\begin{equation}
k^{+}(\zeta_1^{-1/2}u)=e_A(u) f_D(\zeta_1^{-1}u).
\end{equation}
It is clear that terms containing $:\exp(B(z)):$ and $:\exp(C(z)):$ will vanish after taking the difference. Terms containing $:\exp(A(z)):$ and $:\exp(D(z)):$ have nontrivial effects in recursion relations. And $:\exp(A(z)):$ can be considered as the main building block in the construction of $\Phi(u)$. This fact corresponds to the operator $\varphi^{+}(z)$ appearing in the recursion relation (\ref{re:int1F}). In our free field realization, $\varphi^{+}(z)$ is related to $:\exp(A(z)):$ and $:\exp(D(z)):$ while $\varphi^{-}(z)$ is related to $:\exp(B(z)):$ and $:\exp(C(z)):$. 
\par
For general $n$, the situations are similar. We consider the recursion relations (\ref{re:int1E}) and (\ref{re:int1F}) first, and specify recursion relations for $t_n(u)$. It can be checked that
\begin{equation}
:\exp(B(z)):\tilde{\Phi}_{n}(u) -S_1(q_1^{-n}z/u)\tilde{\Phi}_{n}(u) :\exp(B(z)):=0,
\end{equation}
and 
\begin{equation}
:\exp(C(z)):\tilde{\Phi}_{n}(u) -\tilde{\Phi}_{n}(u) :\exp(C(z)):=0.
\end{equation}
So let us focus on terms containing $:\exp(A(z)):$ and $:\exp(D(z)):$. The only distinction is that, we should take the zero mode part into account at this time. The OPE relations of $\tilde{\Phi}_{n}(u)$ and $:\exp(A(z)):$ are given by
\begin{align}
:\exp(A(z)): \tilde{\Phi}_{n}(u) =&\frac{\zeta_1^n z^{-2n}(q_1^{n+1}q_2u/z;q_1)_{\infty}(q_1^{n+1}q_3u/z;q_1)_{\infty}}{1-q_1^{n+1}u/z}
\\ \notag
&\times :\exp(A(z)) \tilde{\Phi}_{n}(u):,
\end{align}
\begin{equation}
\tilde{\Phi}_{n}(u) :\exp(A(z)): = \frac{\zeta_1^n q_1^{-(n+1)n} u^{-2n}(1-q_1^{-n}z/u)}{(q_1^{-n}q_2z/u;q_1)_{\infty}(q_1^{-n}q_3z/u;q_1)_{\infty}} :\exp(A(z)) \tilde{\Phi}_{n}(u): .
\end{equation}
Then 
\begin{align}
&S_1(q^{-n}z/u) \tilde{\Phi}_{n}(u) :\exp(A(z)): 
\\ \notag
&= \frac{\zeta_1^n q_1^{-(n+1)n} u^{-2n}}{(1-q_1^{-n-1}z/u)(q_1^{-n+1}q_2z/u;q_1)_{\infty}(q_1^{-n+1}q_3z/u;q_1)_{\infty}} :\exp(A(z)) \tilde{\Phi}_{n}(u): .
\end{align}
Due to the property (\ref{prop:theta}) of the theta function, we find that 
\begin{align}
&(-q_1^{n+1}u/z) q_1^{-n(n+1)} (z/u)^{2n} \theta_{q_1}(q_1^{n+1}q_2u/z)^{-1}\theta_{q_1}(q_1^{n+1}q_3u/z)^{-1}
\\ \notag
&= (-q_1u/z) \theta_{q_1}(q_1q_2u/z)^{-1}\theta_{q_1}(q_1q_3u/z)^{-1}.
\end{align}
With $e_A'(z)=(-q_1u/z) \theta_{q_1}(q_1q_2u/z)^{-1}\theta_{q_1}(q_1q_3u/z)^{-1}e_A(z)$, we obtain  
\begin{align} 
& \frac{1}{z} (e_A'(z):\exp(A(z)):  \tilde{\Phi}_{n}(u) - S_1(q_1^{-n}z/u) \tilde{\Phi}_{n}(u) e_A(z):\exp(A(z)):)  
\\ \notag
& =  - \frac{\zeta_1^{n}q_1^{-(n+1)^2}u^{-2n-1} e_A(q_1^{n+1}u)} {(q_1^2q_2;q_1)_{\infty}(q_1^2q_3;q_1)_{\infty}} \delta(q_1^{n+1}u/z) \tilde{\Phi}_{n+1}(u) .
\end{align}
Combining this equality with (\ref{re:int1E}), we get the recursion relation of $t_n(u)$:
\begin{equation}
t_{n+1}(u)=-\mathcal{E}^{-1} \frac{\zeta_1^{n}q_1^{-(n+1)^2}u^{-2n-1} e_A(q_1^{n+1}u)} {(q_1^2q_2;q_1)_{\infty}(q_1^2q_3;q_1)_{\infty}} t_n(u).
\end{equation}
Similarly, we have
\begin{align}
&:\exp(D(z)): \tilde{\Phi}_{n}(u) 
\\ \notag
&= \frac{\zeta_1^n z^{2n}:\tilde{\Phi}_{n}(u)\exp(D(z)):}{(1-\zeta_1^{-1}q_1^{n}u/z)(\zeta_1^{-1}q_1^{n+2}q_2u/z;q_1)_{\infty}(\zeta_1^{-1}q_1^{n+2}q_3u/z;q_1)_{\infty}} ,
\end{align}
\begin{align}
\tilde{\Phi}_{n}(u) :\exp(D(z)):=& \frac{\zeta_1^{-n}q_1^{n(n+1)}u^{2n}(\zeta_1q_1^{-n}q_2z/u;q_1)_{\infty}(\zeta_1q_1^{-n}q_3z/u;q_1)_{\infty}}{1-\zeta_1q_1^{-n}z/u} 
\\ \notag
&\times:\tilde{\Phi}_{n}(u)\exp(D(z)):.
\end{align}
With $f_D'(z)=(-\zeta_1^{-1}u/z)\theta_{q_1}(\zeta_1^{-1}q_1^2q_2u/z)\theta_{q_1}(\zeta_1^{-1}q_1^2q_3u/z)f_D(z)$, we obtain
\begin{align} 
& \frac{1}{z} (f_D'(z):\exp(A(z)):  \tilde{\Phi}_{n}(u) - \tilde{\Phi}_{n}(u) f_D(z):\exp(A(z)):) 
\\ \notag
&= - \zeta_1^{-n+1} q_1^{n^2} u^{2n-1} e_A(q_1^{n}u)^{-1}(q_2;q_1)_{\infty}(q_3;q_1)_{\infty}\delta(\zeta_1^{-1}q_1^{n}u/z) \tilde{\Phi}_{n-1}(u) 
\\ \notag
&~~~~~~ \times k^{+}(\zeta_1^{-1/2}q_1^{n}u)\varphi^{+}(\zeta_1^{-1/2}q_1^{n}u) .
\end{align}
Combining this equality with (\ref{re:int1E}), we get another recursion relation of $t_n(u)$:
\begin{equation}
t_{n-1}(u)=-\mathcal{F}^{-1} \zeta_1^{-n+1}q_1^{n^2}u^{2n-1} e_A(q_1^{n}u)^{-1} (q_2;q_1)_{\infty}(q_3;q_1)_{\infty} t_n(u).
\end{equation}
These two recursion relations are compatible since
\begin{equation} 
\mathcal{E} \mathcal{F}=(1-q_2)(1-q_3)(1-q_2^{-1})(1-q_3^{-1}).
\end{equation}
If we set 
\begin{equation}
\mathcal{E}=-(q_1^2q_2;q_1)_{\infty}^{-1}(q_1^2q_3;q_1)_{\infty}^{-1}, \quad
\mathcal{F}=-(q_2;q_1)_{\infty}(q_3;q_1)_{\infty} , \quad
t_0(u)=1,
\end{equation}
the explicit expression of $t_n(u)$ is given by
\begin{align}
t_n(u)=\zeta_1^{n(n+1)/2} q_1^{n(n+1)(2n+1)/6} u^{-n^2} \prod_{j=1}^{n} e_A(q_1^{j}u) ,\qquad  n>0
\\
t_n(u)=\zeta_1^{n(n-1)/2} q_1^{-n(-n+1)(-2n+1)/6} u^{-n^2} \prod_{j=n}^{-1} e_A(q_1^{j}u)^{-1}
,\qquad  n<0.
\end{align}
We have shown that $\Phi_{n}=t_n(u)\tilde{\Phi}_{n}(u)$ satisfies the intertwining relations (\ref{re:int1E}) and (\ref{re:int1F}) with conditions
\begin{equation}
e_A'(z)=(-q_1u/z) \theta_{q_1}(q_1q_2u/z)^{-1}\theta_{q_1}(q_1q_3u/z)^{-1}e_A(z),
\end{equation}
\begin{equation}
e_B'(z)=e_B(z),
\end{equation}
\begin{equation}
f_C'(z)=f_C(z),
\end{equation}
\begin{equation}
f_D'(z)=(-\zeta_1^{-1}u/z)\theta_{q_1}(\zeta_1^{-1}q_1^2q_2u/z)\theta_{q_1}(\zeta_1^{-1}q_1^2q_3u/z)f_D(z).
\end{equation}
The relative shift of factors $k^{\pm,\prime}(z)$ and $k^{\pm}(z)$ can be derived from the above conditions:
\begin{equation}
k^{-,\prime}(z)=e_B'(\zeta_1^{-/2}z) f_C'(\zeta_1^{1/2}z)=e_B(\zeta_1^{-/2}z) f_C(\zeta_1^{1/2}z)=k^{-}(z),
\end{equation}
\begin{equation}
k^{+,,\prime}(z)=e_A'(\zeta_1^{1/2}z) f_D'(\zeta_1^{-1/2}z)=e_A(\zeta_1^{1/2}z) f_D(\zeta_1^{-1/2}z)=k^{+}(z).
\end{equation}
Then the rest intertwining relations (\ref{re:int1Kn}) and (\ref{re:int1Kp}) follow from the following OPE relations:
\begin{equation}
\varphi^{+}(z) \tilde{\Phi}_{n}(u) =S_1(\zeta_1^{-1/2}q_1^{n+1}u/z) :\varphi^{+}(z)\tilde{\Phi}_{n}(u): ,
\end{equation}
\begin{equation}
\tilde{\Phi}_{n}(u)\varphi^{+}(z)=:\varphi^{+}(z)\tilde{\Phi}_{n}(u):
\end{equation}
and
\begin{equation}
\varphi^{-}(z) \tilde{\Phi}_{n}(u)=:\varphi^{-}(z) \tilde{\Phi}_{n}(u):,
\end{equation}
\begin{equation}
\tilde{\Phi}_{n}(u) \varphi^{-}(z)=S_1(\zeta_1^{-1/2} q_1^{-n} z/u)^{-1} :\varphi^{-}(z) \tilde{\Phi}_{n}(u): .
\end{equation}
\subsubsection{Discussions about the zero mode part}
We have chosen a trivial zero mode part for $\Phi_{0}(u)$ above. Actually, there are some degrees of freedom on choosing the zero mode part. If the zero mode part of $\Phi_{0}(u)$ changes, the zero mode part of general $\Phi_{n}(u)$ will change as the same as $\Phi_{0}(u)$ due to the recursion relation (\ref{re:recursionA}). Assume that there is another $\Phi_{0}'(u)$, which is equal to $\Phi_{0}(u)$ up to the zero mode part and that the OPE relations of $\Phi_{0}'(u)$ with $:\exp(A_0(z)):$ and $:\exp(B_0(z)):$ are 
\begin{equation}
:\exp(A_0(z)): \Phi_{0}'(u) =h_A(z;u) :\exp(A_0(z)) \Phi_{0}'(u): ,
\end{equation}
\begin{equation}
\Phi_{0}'(u) :\exp(A_0(z)): =k_A(u) :\exp(A_0(z)) \Phi_{0}'(u): ,
\end{equation}
\begin{equation}
:\exp(B_0(z)): \Phi_{0}'(u) =h_B(z;u) :\exp(B_0(z)) \Phi_{0}'(u): ,
\end{equation}
\begin{equation}
\Phi_{0}'(u) :\exp(B_0(z)): =k_B(u) :\exp(B_0(z)) \Phi_{0}'(u): .
\end{equation}
The OPE relations of $\Phi_{0}'(u)$ with $:\exp(C_0(z)):$ and $:\exp(D_0(z)):$ can be derived from the above OPE relations: 
\begin{equation}
:\exp(D_0(z)): \Phi_{0}'(u) =h_A(\zeta_1z;u)^{-1} :\exp(D_0(z)) \Phi_{0}'(u): ,
\end{equation}
\begin{equation}
\Phi_{0}'(u) :\exp(D_0(z)): =k_A(u)^{-1} :\exp(D_0(z)) \Phi_{0}'(u): ,
\end{equation}
\begin{equation}
:\exp(C_0(z)): \Phi_{0}'(u) =h_B(\zeta_1^{-1}z;u)^{-1} :\exp(C_0(z)) \Phi_{0}'(u): ,
\end{equation}
\begin{equation}
\Phi_{0}'(u) :\exp(C_0(z)): =k_B(u)^{-1} :\exp(C_0(z)) \Phi_{0}'(u): 
\end{equation} 
since
\begin{equation}
D_0(z)=-A_0(\zeta_1z) ,
\end{equation}
\begin{equation}
C_0(z)=-B_0(\zeta_1^{-1}z).
\end{equation}
The relative conditions between $(V,\rho)$ and $(V',\rho')$ become
\begin{equation}
h_A(z;u) e_A'(z)=(-q_1u/z) k_A(u)\theta_{q_1}(q_1q_2u/z)^{-1}\theta_{q_1}(q_1q_3u/z)^{-1}e_A(z),
\end{equation}
\begin{equation}
h_B(z;u)e_B'(z)=k_B(u)e_B(z),
\end{equation}
\begin{equation}
h_B(\zeta_1^{-1}z;u)^{-1} f_C'(z)=k_A(u)^{-1} f_C(z),
\end{equation}
\begin{equation}
h_A(\zeta_1z;u)^{-1} f_D'(z)=(-\zeta_1^{-1}u/z) k_A(u)^{-1} \theta_{q_1}(\zeta_1^{-1}q_1^2q_2u/z)\theta_{q_1}(\zeta_1^{-1}q_1^2q_3u/z)f_D(z).
\end{equation}
And the recursion relation of $t_n(u)$ becomes
\begin{equation}
t_{n+1}(u)=-k_A(u)\mathcal{E}^{-1} \frac{\zeta_1^{n}q_1^{-(n+1)^2}u^{-2n-1} e_A(q_1^{n+1}u)} {(q_1^2q_2;q_1)_{\infty}(q_1^2q_3;q_1)_{\infty}} t_n(u).
\end{equation}
\par
If we choose 
\begin{equation}
\Phi_{0}'(u)= \Phi_{0}(u)\exp(\frac{1}{2}(p_c^2-p_b^1)(\ln q_1u-\frac{1}{2}\ln \zeta_1)+\frac{1}{2}(q_b^1-q_c^2)),
\end{equation}
then the factors $-q_1u/z$ and $-\zeta_1^{-1}u/z$ in the relative conditions will be absorbed.
Then the relative conditions between $(V,\rho)$ and $(V',\rho')$ become
\begin{equation}
e_A'(z)=\theta_{q_1}(q_1q_2u/z)^{-1}\theta_{q_1}(q_1q_3u/z)^{-1}e_A(z),
\end{equation}
\begin{equation}
e_B'(z)=e_B(z),
\end{equation}
\begin{equation}
f_C'(z)=f_C(z),
\end{equation}
\begin{equation}
f_D'(z)=\theta_{q_1}(\zeta_1^{-1}q_1q_2u/z)\theta_{q_1}(\zeta_1^{-1}q_1q_3u/z)f_D(z),
\end{equation}
and the recursion relation of $t_n(u)$ becomes
\begin{equation}
t_{n+1}(u)=-\mathcal{E}^{-1} \frac{\zeta_1^{n+\frac{1}{2}}q_1^{-(n+1)^2-1}u^{-2n-2} e_A(q_1^{n+1}u)} {(q_1^2q_2;q_1)_{\infty}(q_1^2q_3;q_1)_{\infty}} t_n(u).
\end{equation}
\subsubsection{More discussions about the relative conditions}
Let us examine about the relative conditions between $(V,\rho)$ and $(V',\rho')$ in detail. Checking the computation of the relative shift between $e_A(z)$ and $e_A'(z)$ carefully, we may find that the expressions of $(-q_1u/z) \theta_{q_1}(q_1q_2u/z)^{-1}\theta_{q_1}(q_1q_3u/z)^{-1}$ are slightly different for different $n$:
\begin{align}
& (-q_1u/z) \theta_{q_1}(q_1q_2u/z)^{-1}\theta_{q_1}(q_1q_3u/z)^{-1}
\\ \notag 
&=  \frac{(-q_1^{n+1}u/z) q_1^{-n(n+1)} (z/u)^{2n}}{(q_1^{n+1}q_2u/z;q_1)_{\infty}(q_1^{n+1}q_3u/z;q_1)_{\infty}(q_1^{-n}q_2z/u;q_1)_{\infty}(q_1^{-n}q_3z/u;q_1)_{\infty}},
\end{align}
which can be regarded as different Laurent expansions in different annulus 
\begin{equation}
R_n=\{ z|\ |q_1^{n+1}q_2u|<|z|<|q_1^{n}q_2u|,|q_1^{n+1}q_3u|<|z|<|q_1^{n}q_3u| \}.
\end{equation}
If we need to specify this function to be expanded in $R_n$, there would be some unexpected delta functions of form
\begin{equation}
\delta(q_1^{n+m+1}q_2u/z), \quad  \delta(q_1^{n+m+1}q_3u/z),   \qquad  m\geq 0   
\end{equation} 
in the computations of recursion relation (\ref{re:int1E}). To eliminate these unexpected delta functions, $e_A(z)$ is required to contain factors 
\begin{equation}
(q_1^{n+m+1}q_2u/z;q_1)_{\infty}(q_1^{n+m+1}q_3u/z;q_1)_{\infty}.
\end{equation} 
For $m\in\mathbb{Z}$, vertex operator representations $(V,\rho)$ and $(V',\rho')$ with the following factors:
\begin{equation}
e_A(z)=(q_1^{m+1}q_2u/z;q_1)_{\infty}(q_1^{m+1}q_3u/z;q_1)_{\infty},
\end{equation}
\begin{equation}
e_A'(z)=-q_1^{-(m-1)(m+1)} (z/u)^{2m-1}(q_1^{m+1}q_2z/u;q_1)_{\infty}^{-1}(q_1^{m+1}q_3z/u;q_1)_{\infty}^{-1},
\end{equation}
and
\begin{equation}
f_D(z)=(\zeta_1^{-1}q_1^{m+1}q_2u/z;q_1)_{\infty}^{-1}(\zeta_1^{-1}q_1^{m+1}q_3u/z;q_1)_{\infty}^{-1},
\end{equation}
\begin{align}
f_D'(z)=&-\zeta_1^{-2m+1}q_1^{(m-1)(m+1)} (u/z)^{2m-1}
\\ \notag
&\times(\zeta_1q_1^{m+1}q_2z/u;q_1)_{\infty}(\zeta_1q_1^{m+1}q_3z/u;q_1)_{\infty}, 
\end{align}
as well as
\begin{equation}
e_B(z)=e_B'(z)= f_C(z)= f_C'(z)=1,
\end{equation}
are admissible for $\Phi_{n}(u)$.

\subsection{Dual intertwiner}
There exists another type of trivalent intertwiner $\Phi^{*}(u): V \rightarrow V'\otimes V_1(u)$, defined by the intertwining condition
\begin{equation}
\Delta(a) \Phi^{*}(u) = \Phi^{*}(u) a, \qquad \forall a\in \mathcal{U}. 
\end{equation}
Introduce the $n$-component $\Phi_{n}^{*}(u)$ satisfying
\begin{align}
\Phi^{*}(u) (v) = \sum_{n\in\mathbb{Z}} \Phi_n^{*}(u)(v) \otimes [u]_n, \qquad \forall v\in V.
\end{align}
Applying $C_1=\zeta_1$ and $C_2=1$ to the defining relation of the coproduct (\ref{re:coE})-(\ref{re:coKp}), we get 
\begin{equation}
\Phi_{n}^{*}(u) \rho(E(z)) = \rho'(E(z)) \Phi_{n}^{*}(u) +  \rho'(K^{-}(\zeta_1^{1/2}z)) \Phi_{n}^{*}(u) \langle n | E(\zeta_1z) |n-1 \rangle,
\end{equation}
\begin{equation}
\Phi_{n}^{*}(u)\rho(F(z)) = \rho'(F(z)) \Phi_{n}^{*}(u) \langle n |K^{+}(z)|n \rangle + \Phi_{n}^{*}(u) \langle n |F(z)|n+1 \rangle,
\end{equation}
\begin{equation}
\Phi_{n}^{*}(u) \rho(K^{+}(z)) = \rho'(K^{+}(z))\Phi_{n}^{*}(u) \langle n | K^{+}(\zeta_1^{-1/2}z) |n \rangle ,
\end{equation}
\begin{equation}
\Phi_{n}^{*}(u) \rho(K^{-}(z)) = \rho'(K^{-}(z)) \Phi_{n}^{*}(u)  \langle n |K^{-}(\zeta_1^{1/2}z)|n \rangle  .
\end{equation}
More precisely, we have 
\begin{align} \label{re:int2E}
& \frac{1}{z} (e_A'(z):\exp(A(z)): +e_B'(z):\exp(B(z)):) \Phi_{n}^{*}(u) 
\\ \notag
&- \frac{1}{z} \Phi_{n}^{*}(u) (e_A(z):\exp(A(z)): +e_B(z):\exp(B(z)):) 
\\ \notag
& = - \mathcal{E} \delta(\zeta_1^{-1}q_1^{n}u/z) k^{-,\prime}(\zeta_1^{1/2}z) \varphi^{-,\prime}(\zeta_1^{1/2}z) \Phi_{n-1}^{*}(u),
\end{align}
\begin{align} 
& \frac{1}{z} S_1(q_1^{n+1} u/z)(f_C'(z):\exp(C(z)):+ f_D'(z):\exp(D(z)):) \Phi_{n}^{*}(u)
\\ \notag
& - \frac{1}{z} \Phi_{n}^{*}(u) (f_C(z):\exp(C(z)):+ f_D(z):\exp(D(z)):)
\\ \notag
&= - \mathcal{F} \delta(q_1^{n+1}u/z) \Phi_{n+1}^{*}(u) ,
\end{align}
\begin{equation} 
k^{+}(z) \Phi_{n}^{*}(u) \varphi^{+}(z) = S_1(\zeta_1^{1/2} q_1^{n+1} u/z) k^{+,\prime}(z) \varphi^{+}(z) \Phi_{n}^{*}(u),
\end{equation}
\begin{equation} 
k^{-}(z) \Phi_{n}^{*}(u) \varphi^{-}(z) = S_1(\zeta_1^{1/2} q_1^{-n} z/u) k^{-,\prime}(z) \varphi^{-}(z) \Phi_{n}^{*}(u).
\end{equation}
We find that $\varphi^{-}(z)$ appears in recursion relation (\ref{re:int2E}) in contrary to the case of $\Phi(u)$. At this time, it is $:\exp(C(z)):$ serves as the main building block in the construction of $\Phi^{*}(u)$. Terms containing $:\exp(A(z)):$ and $:\exp(D(z)):$ will be canceled in the recursion relations. Define
\begin{align}
\tilde{\Phi}_n^{*}(u) = &\exp\left(-\sum_{r=1}\frac{q_1^{r(n+1)}}{1-q_1^r}(a_{-r}^2+c_{-r}^1)u^r\right)
\\ \notag
&\times \exp((p_b^2-p_c^1)(\ln q_1^{\frac{n(n+1)}{2}}u^n-\frac{n}{2}\ln\zeta_1)+n(q_c^1-q_b^2))
\\ \notag
&\times \exp\left(\sum_{r=1}\frac{q_1^{-nr}}{1-q^r}(-\zeta_1^rb_r^2+c_r^1)u^{-r}\right),
\end{align}
and 
\begin{equation}
\Phi_n^{*}(u)=t_n^{*}(u)\tilde{\Phi}_n^{*}(u).
\end{equation}
The relative conditions between $(V,\rho)$ and $(V',\rho')$ are 
\begin{equation}
e_A'(z)=e_A(z),
\end{equation}
\begin{equation}
e_B'(z)=(-\zeta_1^{-1}u/z)\theta_{q_1}(\zeta_1^{-1}q_1^2q_2u/z)\theta_{q_1}(\zeta_1^{-1}q_1^2q_3u/z)e_B(z),
\end{equation}
\begin{equation}
f_C'(z)=(-q_1u/z)\theta_{q_1}(q_1q_2u/z)^{-1}\theta_{q_1}(q_1q_3u/z)^{-1}f_C(z),
\end{equation}
\begin{equation}
f_D'(z)=f_D(z).
\end{equation}
And the recursion relation of $t_n^{*}(u)$ is given by 
\begin{equation}
t_{n+1}^{*}(u)=\mathcal{F}^{-1} \frac{\zeta_1^{n}q_1^{-(n+1)^2}u^{-2n-1} f_C(q_1^{n+1}u)} {(q_1^2q_2;q_1)_{\infty}(q_1^2q_3;q_1)_{\infty}} t_n^{*}(u).
\end{equation}
There are also some degrees of freedom to choosing the zero mode part. 

\section{Conclusion and outlook}
In this paper, we have constructed a unified free field realization of the Ding-Iohara algebra at arbitrary levels, satisfying a set of generalized Serre relations. Different factorizations of the structure function $g(z)$ are adopted in our construction and the previous free field realization of horizontal representations. In principle, this difference can be considered as some twists, as we have seen in the representations with level $(\gamma,1)$. The free field horizontal representations have many applications via intertwiners in the gauge theory and string theory. We have constructed vector intertwiners by use of this new free field realization, which generalize the original vector intertwiners. We expect that our construction will also have its role in these areas. The constructions of Fock intertwiners and MacMahon intertwiners are in progress. There are other attractive objects in integrable system, like the R-matrix and screening charges. Using this new free field realization might lead to some novel discoveries in the study of these objects. Furthermore, the concept of the quantum toroidal algebra of $\mathfrak{gl}_1$ is extended to quiver quantum toroidal algebras in \cite{NW}. Our construction can potentially be extended to quiver quantum toroidal algebras, albeit with the Serre relations likely requiring analogous generalizations.

\acknowledgments
One of the authors (Ding) gratefully acknowledges financial support from the National Natural Science Foundation of China (Grant No. 11775299). The authors are grateful to the anonymous referee for useful comments.

\section*{Declarations}

\begin{itemize}
\item The authors declare that they have no known competing financial interests.
\item This study is a theoretical analysis and does not generate new experimental data.  
\end{itemize}

\appendix
\section{Serre relations in horizontal representations}
Substituting (\ref{def:eta}) into (\ref{re:Serre_E}), 
we find that the OPE factor is given by
\begin{align}
h(&z_1,z_2,z_3)
\\ \notag
=&\left(\frac{z_2}{z_3}+\frac{z_2}{z_1}-\frac{z_3}{z_2}-\frac{z_1}{z_2}\right) (S_3(z_3/z_2)^{-1}S_3(z_2/z_1)^{-1}S_3(z_3/z_1)^{-1}
\\ \notag
&+S_3(z_1/z_2)^{-1}S_3(z_3/z_1)^{-1}S_3(z_3/z_2)^{-1})
\\ \notag
&+\left(\frac{z_1}{z_2}+\frac{z_1}{z_3}-\frac{z_2}{z_1}-\frac{z_3}{z_1}\right) 
(S_3(z_1/z_2)^{-1}S_3(z_3/z_1)^{-1}S_3(z_3/z_2)^{-1}
\\ \notag
&+S_3(z_1/z_3)^{-1}S_3(z_2/z_1)^{-1}S_3(z_2/z_3)^{-1})
\\ \notag
&+\left(\frac{z_3}{z_1}+\frac{z_3}{z_2}-\frac{z_1}{z_3}-\frac{z_2}{z_3}\right)
(S_3(z_3/z_1)^{-1}S_3(z_2/z_3)^{-1}S_3(z_2/z_1)^{-1}
\\ \notag
&+S_3(z_3/z_2)^{-1}S_3(z_1/z_3)^{-1}S_3(z_1/z_2)^{-1}).
\end{align}
Denote
\begin{equation*}
S(z,w)=\frac{(z-w)(z-q_3^{-1}w)}{(z-q_1w)(z-q_2w)},
\end{equation*}
as a rational function. Assuming that $f(z_1,\dots,z_n)$ is a rational function, we denote 
\begin{equation*}
( f(z_1,\dots,z_n) )_{z_{i_1},\dots,z_{i_n}} ,
\end{equation*}
as the Laurent expansion in the region $|z_{i_1}|\gg\dots\gg|z_{i_n}|$. In this notation, 
\begin{equation*}
(S(z,w))_{z,w}=S_3(w/z)^{-1}.
\end{equation*}

We classify the terms in $h(z_1,z_2,z_3)$ into two sets. One set contains $(S(z_1,z_2))_{z_1,z_2}$ which is a Laurent expansion in the region $|z_1|\gg|z_2|$ while the other contains $(S(z_1,z_2))_{z_2,z_1}$ which is a Laurent expansion in the region $|z_2|\gg|z_1|$.
First, consider those containing $(S(z_1,z_2))_{z_2,z_1}$. To compute the sum of these three terms, we need to make them to be Laurent expansions in the same region $|z_1|\gg|z_2|\gg|z_3|$.  The following identities are used: 
\begin{align*}
\Delta_{S}(u/z)
&:=(S(z,w))_{z,w}-(S(z,w))_{w,z}
\\
&=
\frac{(1-q_1^{-1})(1-q_2)}{(1-q_1^{-1}q_2)}\delta(q_1w/z)+\frac{(1-q_2^{-1})(1-q_1)}{(1-q_1q_2^{-1})}\delta(q_2w/z),
\end{align*}
and 
\begin{align*}
(S&(z,w)S(z,u))_{z,w,u}-(S(z,w)S(z,u))_{w,u,z}
\\
=
&(S(z,w))_{w,z}\Delta_{S}(u/z)+(S(z,u))_{z,u}\Delta_{S}(w/z).
\end{align*}
Denote
\begin{equation*}
S_{123}(z_1,z_2,z_3)=\left(\frac{z_2}{z_3}+\frac{z_2}{z_1}-\frac{z_3}{z_2}-\frac{z_1}{z_2}\right)
S(z_1,z_2)S(z_2,z_3)S(z_1,z_3),
\end{equation*}
\begin{equation*}
S_{132}(z_1,z_2,z_3)=\left(\frac{z_3}{z_1}+\frac{z_3}{z_2}-\frac{z_1}{z_3}-\frac{z_2}{z_3}\right)
S(z_1,z_3)S(z_3,z_2)S(z_1,z_2),
\end{equation*}
\begin{equation*}
S_{312}(z_1,z_2,z_3)=\left(\frac{z_1}{z_2}+\frac{z_1}{z_3}-\frac{z_2}{z_1}-\frac{z_3}{z_1}\right)
S(z_3,z_1)S(z_1,z_2)S(z_3,z_2),
\end{equation*}
as rational functions. Then  
\begin{align} \label{eq:sumS12}
(S&_{123}(z_1,z_2,z_3))_{z_1,z_2,z_3}+(S_{132}(z_1,z_2,z_3))_{z_1,z_3,z_2}+(S_{312}(z_1,z_2,z_3))_{z_3,z_1,z_2}
\\ \notag
=
&(S_{123}(z_1,z_2,z_3))_{z_1,z_2,z_3}+(S_{132}(z_1,z_2,z_3))_{z_1,z_2,z_3}+(S_{312}(z_1,z_2,z_3))_{z_1,z_2,z_3}
\\ \notag
&+(K_{z_3=q_1z_1}(z_1,z_2))_{z_1,z_2}\delta(q_1z_1/z_3)+(K_{z_3=q_2z_1}(z_1,z_2))_{z_1,z_2}\delta(q_2z_1/z_3)
\\ \notag
&+(K_{z_3=q_1z_2}(z_1,z_2))_{z_1,z_2}\delta(q_1z_2/z_3)+(K_{z_3=q_2z_2}(z_1,z_2))_{z_1,z_2}\delta(q_2z_2/z_3),
\end{align}
where
\begin{equation*}
K_{z_3=q_1z_1}(z_1,z_2)=\frac{(-1+q_1)(-1+q_2)(q_1z_1-z_2)(q_1z_1+z_2)(-z_1+q_1q_2z_2)}{q_1(q_1-q_2)z_1z_2(q_1z_1-q_2z_2)},
\end{equation*}
\begin{equation*}
K_{z_3=q_2z_1}(z_1,z_2)=\frac{(-1+q_1)(-1+q_2)(q_2z_1-z_2)(q_2z_1+z_2)(-z_1+q_1q_2z_2)}{q_2(q_1-q_2)z_1z_2(-q_2z_1+q_1z_2)},
\end{equation*}
\begin{equation*}
K_{z_3=q_1z_2}(z_1,z_2)=-\frac{(-1+q_1)(-1+q_2)(q_1q_2z_1-z_2)(-z_1+q_1z_2)(z_1+q_1z_2)}{q_1(q_1-q_2)z_1z_2(-q_2z_1+q_1z_2)},
\end{equation*}
\begin{equation*}
K_{z_3=q_2z_2}(z_1,z_2)=-\frac{(-1+q_1)(-1+q_2)(q_1q_2z_1-z_2)(-z_1+q_2z_2)(z_1+q_2z_2)}{q_2(q_1-q_2)z_1z_2(q_1z_1-q_2z_2)}.
\end{equation*}
Each of them contains only one simple pole. The sum in the second line of (\ref{eq:sumS12}) gives the Laurent expansion of the following function in the region $|z_1|,|z_2|\gg|z_3|$:
\begin{align}\label{eq:S12}
&\frac{(1-q_1)(1-q_2)(1-q_1q_2)z_3(q_1q_2z_1-z_2)(-z_1+q_1q_2z_2)}{z_1z_2(z_1-q_1z_3)(z_1-q_1^{-1}z_3)(z_2-q_1z_3)(z_2-q_1^{-1}z_3)}
\\ \notag
&\times 
\frac{(z_1+z_3)(z_2+z_3)(z_1z_2-z_3^2)\prod_{i<j}(z_i-z_j)}{(z_1-q_2z_3)(z_1-q_2^{-1}z_3)(z_2-q_2z_3)(z_2-q_2^{-1}z_3)},
\end{align}
since the poles like $z_2=cz_1$ vanish.
Similarly, denote
\begin{equation*}
S_{321}(z_1,z_2,z_3)=\left(\frac{z_2}{z_3}+\frac{z_2}{z_1}-\frac{z_3}{z_2}-\frac{z_1}{z_2}\right)
S(z_3,z_2)S(z_2,z_1)S(z_3,z_1),
\end{equation*}
\begin{equation*}
S_{231}(z_1,z_2,z_3)=\left(\frac{z_3}{z_1}+\frac{z_3}{z_2}-\frac{z_1}{z_3}-\frac{z_2}{z_3}\right)
S(z_2,z_3)S(z_3,z_1)S(z_2,z_1),
\end{equation*}
\begin{equation*}
S_{213}(z_1,z_2,z_3)=\left(\frac{z_1}{z_2}+\frac{z_1}{z_3}-\frac{z_2}{z_1}-\frac{z_3}{z_1}\right)
S(z_2,z_1)S(z_1,z_3)S(z_2,z_3),
\end{equation*}
as rational functions. Then make them to be Laurent expansions in the same region $|z_2|\gg|z_1|\gg|z_3|$:
\begin{align} \label{eq:sumS21}
&(S_{321}(z_1,z_2,z_3))_{z_3,z_2,z_1}+(S_{231}(z_1,z_2,z_3))_{z_2,z_3,z_1}+(S_{213}(z_1,z_2,z_3))_{z_2,z_1,z_3}
\\ \notag
=
&(S_{321}(z_1,z_2,z_3))_{z_2,z_1,z_3}+(S_{231}(z_1,z_2,z_3))_{z_2,z_1,z_3}+(S_{213}(z_1,z_2,z_3))_{z_2,z_1,z_3}
\\ \notag
&-(K_{z_3=q_1z_1}(z_1,z_2))_{z_2,z_1}\delta(q_1z_1/z_3)-(K_{z_3=q_2z_1}(z_1,z_2))_{z_2,z_1}\delta(q_2z_1/z_3)
\\ \notag
&-(K_{z_3=q_1z_2}(z_1,z_2))_{z_2,z_1}\delta(q_1z_2/z_3)-(K_{z_3=q_2z_2}(z_1,z_2))_{z_2,z_1}\delta(q_2z_2/z_3).
\end{align}
The sum in the second line of (\ref{eq:sumS21}) gives the opposite of (\ref{eq:S12}) while the factors of delta functions are Laurent expansions in the region $|z_1|\gg|z_2|$. Again, delta functions will be produced but they will cancel with each other:
\begin{align*}
&[(K_{z_3=q_1z_1}(z_1,z_2))_{z_1,z_2}-(K_{z_3=q_1z_1}(z_1,z_2))_{z_2,z_1}]\delta(q_1z_1/z_3)
\\
&+[(K_{z_3=q_2z_2}(z_1,z_2))_{z_1,z_2}-(K_{z_3=q_2z_2}(z_1,z_2))_{z_2,z_1}]\delta(q_2z_2/z_3)=0,
\end{align*} 
and 
\begin{align*}
&[(K_{z_3=q_2z_1}(z_1,z_2))_{z_1,z_2}-(K_{z_3=q_2z_1}(z_1,z_2))_{z_2,z_1}]\delta(q_2z_1/z_3)
\\
&+ [(K_{z_3=q_1z_2}(z_1,z_2))_{z_1,z_2}-(K_{z_3=q_1z_2}(z_1,z_2))_{z_2,z_1}]\delta(q_1z_2/z_3)=0.
\end{align*} 

\section{Calculation of $h_{AAA}$}
The strategy is the same as above. Classify the terms in $h_{AAA}(z_1,z_2,z_3)$ into two sets. One set contains 
\begin{equation*}
z_1^{-2}g_{+}(z_2/z_1)^{-1}=\frac{z_1-z_2}{(z_1-q_1z_2)(z_1-q_2z_2)(z_1-q_3z_2)}.
\end{equation*}  
Denote 
\begin{equation*}
G_{123}(z_1,z_2,z_3)=\left(\frac{z_2}{z_3}+\frac{z_2}{z_1}-\frac{z_3}{z_2}-\frac{z_1}{z_2}\right)
g_{+}(z_1,z_2)g_{+}(z_2,z_3)g_{+}(z_1,z_3),
\end{equation*}
\begin{equation*}
G_{132}(z_1,z_2,z_3)=\left(\frac{z_3}{z_1}+\frac{z_3}{z_2}-\frac{z_1}{z_3}-\frac{z_2}{z_3}\right)
g_{+}(z_1,z_3)g_{+}(z_3,z_2)g_{+}(z_1,z_2),
\end{equation*}
\begin{equation*}
G_{312}(z_1,z_2,z_3)=\left(\frac{z_1}{z_2}+\frac{z_1}{z_3}-\frac{z_2}{z_1}-\frac{z_3}{z_1}\right)
g_{+}(z_3,z_1)g_{+}(z_1,z_2)g_{+}(z_3,z_2),
\end{equation*}
as rational functions.
Then 
\begin{align*}
&(G_{123}(z_1,z_2,z_3))_{z_1,z_2,z_3}
\\
&=\left(\frac{z_2}{z_3}+\frac{z_2}{z_1}-\frac{z_3}{z_2}-\frac{z_1}{z_2}\right) z_1^{-4}z_2^{-2}g_{+}(z_3/z_2)^{-1}g_{+}(z_2/z_1)^{-1}g_{+}(z_3/z_1)^{-1},
\end{align*}
\begin{align*}
&(G_{132}(z_1,z_2,z_3))_{z_1,z_3,z_2}
\\
&=\left(\frac{z_3}{z_1}+\frac{z_3}{z_2}-\frac{z_1}{z_3}-\frac{z_2}{z_3}\right) z_1^{-4}z_2^{-2}g_{+}(z_3/z_2)^{-1}g_{+}(z_2/z_1)^{-1}g_{+}(z_3/z_1)^{-1},
\end{align*}
\begin{align*}
&(G_{312}(z_1,z_2,z_3))_{z_3,z_1,z_2}
\\
&=\left(\frac{z_1}{z_2}+\frac{z_1}{z_3}-\frac{z_2}{z_1}-\frac{z_3}{z_1}\right) z_1^{-4}z_2^{-2}g_{+}(z_1/z_3)^{-1}g_{+}(z_2/z_1)^{-1}g_{+}(z_2/z_3)^{-1},
\end{align*}
are the three terms containing $z_1^{-2}g_{+}(z_2/z_1)^{-1}$.
To compute the sum of these three terms, we need to make them to be Laurent expansions in the same region $|z_1|\gg|z_2|\gg|z_3|$. The following equalities are used:
\begin{align*}
\Delta_{+}(w/z) :=
&(g_{+}(z,w))_{z,w}-(g_{+}(z,w))_{w,z}
\\ \notag
=
&\frac{1-q_1^{-1}}{(1-q_1^{-1}q_2)(1-q_1^{-1}q_3)z^2}\delta(q_1w/z)
\\ \notag
&+\frac{1-q_2^{-1}}{(1-q_2^{-1}q_1)(1-q_2^{-1}q_3)z^2}\delta(q_2w/z)
\\ \notag
&+\frac{1-q_3^{-1}}{(1-q_3^{-1}q_1)(1-q_3^{-1}q_1)z^2}\delta(q_3w/z),
\end{align*}
and 
\begin{align*}
&(g_{+}(z,w)g_{+}(z,u))_{z,w,u}-(g_{+}(z,w)g_{+}(z,u))_{w,u,z}
\\
=
&(g_{+}(z,w))_{w,z}\Delta_{+}(u/z)+(g_{+}(z,u))_{z,u}\Delta_{+}(w/z).
\end{align*}
Then we get 
\begin{align} \label{eq:sum12}
(G&_{123}(z_1,z_2,z_3))_{z_1,z_2,z_3}+(G_{132}(z_1,z_2,z_3))_{z_1,z_3,z_2}+(G_{312}(z_1,z_2,z_3))_{z_3,z_1,z_2}
\\ \notag
=&(G_{123}(z_1,z_2,z_3))_{z_1,z_2,z_3}+(G_{132}(z_1,z_2,z_3))_{z_1,z_2,z_3}+(G_{312}(z_1,z_2,z_3))_{z_1,z_2,z_3}
\\ \notag
&+(H_{z_3=q_1z_1}(z_1,z_2))_{z_1,z_2}\delta(q_1z_1/z_3)+(H_{z_3=q_2z_1}(z_1,z_2))_{z_1,z_2}\delta(q_2z_1/z_3)
\\ \notag
&+(H_{z_3=q_3z_1}(z_1,z_2))_{z_1,z_2}\delta(q_3z_1/z_3)+(H_{z_3=q_1z_2}(z_1,z_2))_{z_1,z_2}\delta(q_1z_2/z_3)
\\ \notag
&+(H_{z_3=q_2z_2}(z_1,z_2))_{z_1,z_2}\delta(q_2z_2/z_3)+(H_{z_3=q_3z_2}(z_1,z_2))_{z_1,z_2}\delta(q_3z_2/z_3).
\end{align}
For $i=1,2,3$,
\begin{equation*}
H_{z_3=q_iz_2}(z_1,z_2)
=
\frac{(1-q_i)(q_i^2z_2^2-z_1^2)}{z_1^{1}z_2^{3}\prod_{j\neq i}(1-q_iq_j^{-1})(z_1-q_j^{-1}z_2)(z_1-q_iq_j^{-1}z_2)}
\end{equation*}
is a rational function with four simple poles at $z_2=q_jz_1,q_i^{-1}q_jz_1(j\neq i)$;
\begin{equation*}
H_{z_3=q_iz_1}(z_1,z_2)
=
\frac{(1-q_i)(q_i^2z_1^2-z_2^2)}{z_2^{1}z_1^{3}\prod_{j\neq i}(1-q_iq_j^{-1})(z_2-q_j^{-1}z_1)(z_2-q_iq_j^{-1}z_1)},
\end{equation*}
is a rational function with four simple poles at $z_2=q_j^{-1}z_1,q_iq_j^{-1}z_1(j\neq i)$.

The sum in the second line of (\ref{eq:sum12}) is the Laurent expansion of 
\begin{equation} \label{eq:G12}
\frac{\prod_{i}(1-q_i)\prod_{i<j}(z_i-z_j)z_3(z_1+z_3)(z_2+z_3)(z_1z_2-z_3^2)}{z_1z_2\prod_i (z_1-q_iz_3) (z_3-q_iz_1)  (z_2-q_iz_3) (z_3-q_iz_2)},
\end{equation}
in the region $|z_1|\gg|z_2|\gg|z_3|$. Note that the poles at $z_2=q_i^{-1}z_1$ vanish.

The other set includes
\begin{equation*}
(G_{321}(z_1,z_2,z_3))_{z_3,z_2,z_1},(G_{231}(z_1,z_2,z_3))_{z_2,z_3,z_1},(G_{213}(z_1,z_2,z_3))_{z_1,z_2,z_3},
\end{equation*}
where we denote 
\begin{equation*}
G_{321}(z_1,z_2,z_3)=\left(\frac{z_2}{z_3}+\frac{z_2}{z_1}-\frac{z_3}{z_2}-\frac{z_1}{z_2}\right)
g_{+}(z_3,z_2)g_{+}(z_2,z_1)g_{+}(z_3,z_1),
\end{equation*}
\begin{equation*}
G_{231}(z_1,z_2,z_3)=\left(\frac{z_3}{z_1}+\frac{z_3}{z_2}-\frac{z_1}{z_3}-\frac{z_2}{z_3}\right)
g_{+}(z_2,z_3)g_{+}(z_3,z_1)g_{+}(z_2,z_1),
\end{equation*}
\begin{equation*}
G_{213}(z_1,z_2,z_3)=\left(\frac{z_1}{z_2}+\frac{z_1}{z_3}-\frac{z_2}{z_1}-\frac{z_3}{z_1}\right)
g_{+}(z_2,z_1)g_{+}(z_1,z_3)g_{+}(z_2,z_3),
\end{equation*}
as rational functions. They contain
\begin{equation*}
z_2^{-2}g_{+}(z_1/z_2)^{-1}=\frac{z_2-z_1}{(z_2-q_1z_1)(z_2-q_2z_1)(z_2-q_3z_1)}.
\end{equation*}
Similarly, we  make them to be Laurent expansions in the same region $|z_2|\gg|z_1|\gg|z_3|$:
\begin{align} \label{eq:sum21}
(G&_{321}(z_1,z_2,z_3))_{z_3,z_2,z_1}+(G_{231}(z_1,z_2,z_3))_{z_2,z_3,z_1}+(G_{213}(z_1,z_2,z_3))_{z_1,z_2,z_3}
\\ \notag
=
&(G_{321}(z_1,z_2,z_3))_{z_2,z_1,z_3}+(G_{231}(z_1,z_2,z_3))_{z_2,z_1,z_3}+(G_{213}(z_1,z_2,z_3))_{z_2,z_1,z_3}
\\ \notag
&-(H_{z_3=q_1z_1}(z_1,z_2))_{z_2,z_1}\delta(q_1z_1/z_3)-(H_{z_3=q_2z_1}(z_1,z_2))_{z_2,z_1}\delta(q_2z_1/z_3)
\\ \notag
&-(H_{z_3=q_3z_1}(z_1,z_2))_{z_2,z_1}\delta(q_3z_1/z_3)-(H_{z_3=q_1z_2}(z_1,z_2))_{z_2,z_1}\delta(q_1z_2/z_3)
\\ \notag
&-(H_{z_3=q_2z_2}(z_1,z_2))_{z_2,z_1}\delta(q_2z_2/z_3)-(H_{z_3=q_3z_2}(z_1,z_2))_{z_2,z_1}\delta(q_3z_2/z_3).
\end{align}

The sum in the second line of (\ref{eq:sum21}) is the Laurent expansion of 
\begin{equation} \label{eq:G21}
-\frac{\prod_{i}(1-q_i)\prod_{i<j}(z_i-z_j)z_3(z_1+z_3)(z_2+z_3)(z_1z_2-z_3^2)}{z_1z_2\prod_i (z_1-q_iz_3) (z_3-q_iz_1)  (z_2-q_iz_3) (z_3-q_iz_2)},
\end{equation}
in the region $|z_2|\gg|z_1|\gg|z_3|$. Since the poles at $z_2=q_iz_1$ vanish, it is indeed the Laurent expansion in the region $|z_2|,|z_1|\gg|z_3|$. Thus, (\ref{eq:G12}) and (\ref{eq:G21}) cancel each other. The important thing is that the factors of the delta functions are Laurent expansion in the region $|z_2|\gg|z_1|$ at this time. When combining (\ref{eq:sum12}) and (\ref{eq:sum21}), delta functions in terms of $z_1$ and $z_2$ will be produced, such as 
\begin{align*}
(H&_{z_3=q_1z_1}(z_1,z_2))_{z_1,z_2}-(H_{z_3=q_1z_1}(z_1,z_2))_{z_2,z_1}
\\ 
=
&*\delta(q_2^{-1}z_1/z2)+*\delta(q_1q_2^{-1}z_1/z2)+*\delta(q_1q_2z_1/z2)+*\delta(q_1^2q_2z_1/z2).
\end{align*}
Then 
\begin{align*}
[(&H_{z_3=q_1z_1}(z_1,z_2))_{z_1,z_2}-(H_{z_3=q_1z_1}(z_1,z_2))_{z_2,z_1}]\delta(q_1z_1/z_3)
\\ 
=
&\delta(q_1z_1/z_3)[*\delta(q3^{-1}z_2/z3)+*\delta(q_2z_2/z3)
\\
&~~~~~~~~~~~~~~+*\delta(q_2^{-1}z_2/z3)+*\delta(q_3z_2/z3)],
\end{align*}
where we use equalities like
\begin{equation*}
\delta(q_1z_1/z_3)\delta(q_2^{-1}z_1/z2)=\delta(q_1z_1/z_3)\delta(q_2z_2/z3).
\end{equation*}
Generally, there are two kinds of terms in 
\begin{equation*}
[(H_{z_3=q_iz_1}(z_1,z_2))_{z_1,z_2}-(H_{z_3=q_iz_1}(z_1,z_2))_{z_2,z_1}]\delta(q_iz_1/z_3).
\end{equation*}
One is in the form of
\begin{equation*}
\delta(q_iz_1/z_3)\delta(q_jz_2/z3), \quad i\neq j
\end{equation*}
like
\begin{equation*}
\delta(q_1z_1/z_3)\delta(q_2z_2/z3),\delta(q_1z_1/z_3)\delta(q_3z_2/z3).
\end{equation*}
The other one is in the form of 
\begin{equation*}
\delta(q_iz_1/z_3)\delta(q_j^{-1}z_2/z3), \quad i\neq j
\end{equation*}
like
\begin{equation*}
\delta(q_1z_1/z_3)\delta(q_2^{-1}z_2/z3),\delta(q_1z_1/z_3)\delta(q_3^{-1}z_2/z3).
\end{equation*}
Similarly, there are two kinds of terms 
\begin{equation*}
\delta(q_iz_2/z_3)\delta(q_jz_1/z3), \ \delta(q_iz_2/z_3)\delta(q_j^{-1}z_1/z3), \quad i\neq j
\end{equation*}
in 
\begin{equation*}
[(H_{z_3=q_iz_2}(z_1,z_2))_{z_1,z_2}-(H_{z_3=q_iz_2}(z_1,z_2))_{z_2,z_1}]\delta(q_iz_2/z_3)
\end{equation*}
Terms in the form of
\begin{equation*}
\delta(q_iz_1/z_3)\delta(q_jz_2/z3), \quad i\neq j
\end{equation*}
appear twice and cancel with each other. However, terms in the form of
\begin{equation*}
\delta(q_i^{-1}z_1/z_3)\delta(q_jz_2/z3),\  \delta(q_iz_1/z_3)\delta(q_j^{-1}z_2/z3), \quad i\neq j
\end{equation*}
are left.
Finally, we get
\begin{align} 
h&_{AAA}(z_1,z_2,z_3)
\\ \notag
=
&(q_1-q_2)^{-1}(q_1-q_3)^{-1}(q_2-q_3)^{-1}z_1^{-2}z_2^{-2}z_3^{-2}
\\ \notag
&\times
[\delta(q_1z_2/z_3)\delta(q_3^{-1}z_1/z_3)+\delta(q_2z_2/z_3)\delta(q_1^{-1}z_1/z_3)+\delta(q_3z_2/z_3)\delta(q_2^{-1}z_1/z_3)
\\ \notag
&+\delta(q_1z_1/z_3)\delta(q_3^{-1}z_2/z_3)+\delta(q_2z_1/z_3)\delta(q_1^{-1}z_2/z_3)+\delta(q_3z_1/z_3)\delta(q_2^{-1}z_2/z_3)
\\ \notag
&-\delta(q_1z_2/z_3)\delta(q_2^{-1}z_1/z_3)-\delta(q_2z_2/z_3)\delta(q_3^{-1}z_1/z_3)-\delta(q_3z_2/z_3)\delta(q_1^{-1}z_1/z_3)
\\ \notag
&-\delta(q_1z_1/z_3)\delta(q_2^{-1}z_2/z_3)-\delta(q_2z_1/z_3)\delta(q_3^{-1}z_2/z_3)-\delta(q_3z_1/z_3)\delta(q_1^{-1}z_2/z_3)
].
\end{align}

\section{Calculation of $h_{BAA}$}
We classify the terms in $h_{BAA}(z_1,z_2,z_3)$ into two sets. One set contains 
\begin{equation*}
z_3^{-2}g_{+}(z_2/z_3)^{-1}=\frac{z_3-z_2}{(z_3-q_1z_2)(z_3-q_2z_2)(z_3-q_3z_2)},
\end{equation*}  
including the three terms
\begin{equation*}
\left(\frac{z_2}{z_3}+\frac{z_2}{z_1}-\frac{z_3}{z_2}-\frac{z_1}{z_2}\right)
z_2^{-2}g_{+}(z_3/z_2)^{-1},
\end{equation*}
\begin{equation*}
\left(\frac{z_1}{z_2}+\frac{z_1}{z_3}-\frac{z_2}{z_1}-\frac{z_3}{z_1}\right)
z_2^{-2}
g_{+}(z_3/z_2)^{-1}g(z_1/z_2)^{-1},
\end{equation*}
\begin{equation*}
\left(\frac{z_3}{z_1}+\frac{z_3}{z_2}-\frac{z_1}{z_3}-\frac{z_2}{z_3}\right)
z_2^{-2}g_{+}(z_3/z_2)^{-1}g(z_1/z_2)^{-1}g(z_1/z_3)^{-1}.
\end{equation*}
All these terms can be considered as Laurent series in the region $|z_2|\gg|z_3|\gg|z_1|$. Formally, the sum is 
\begin{equation*}
\frac{(1-q_1)(1-q_2)(1-q_3)z_1(z_2-z_3)(z_1+z_2)(z_1+z_3)(z_1z_2-z_3^2)}{z_2z_3 \prod_i (z_2-q_iz_1) \prod_i (z_3-q_iz_1)}.
\end{equation*}
Note that the poles at $z_3=q_iz_2$ eliminate. It is a Laurent expansion in the region $|z_2|,|z_3|\gg|z_1|$.
The other contains 
\begin{equation*}
z_2^{-2}g_{+}(z_3/z_2)^{-1}=\frac{z_2-z_3}{(z_2-q_1z_3)(z_2-q_2z_3)(z_2-q_3z_3)},
\end{equation*}
of which the sum is 
\begin{equation*}
-\frac{(1-q_1)(1-q_2)(1-q_3)z_1(z_2-z_3)(z_1+z_2)(z_1+z_3)(z_1z_2-z_3^2)}{z_2z_3 \prod_i (z_2-q_iz_1) \prod_i (z_3-q_iz_1)}.
\end{equation*}
It is also a Laurent expansion in the region $|z_2|,|z_3|\gg|z_1|$. Thus
\begin{equation*}
h_{BAA}(z_1,z_2,z_3)=0.
\end{equation*}


\end{document}